\documentclass[fleqn,11pt]{article}
\usepackage{geometry}                
\usepackage{graphicx}
\usepackage{amssymb}
\usepackage{amsmath}
\usepackage{xfrac}
\usepackage{authblk}
\usepackage{setspace}
\usepackage[usenames,dvipsnames]{xcolor}
\usepackage{hyperref}
\usepackage{amsfonts}
\usepackage{pdfpages}
\usepackage{xcolor}
\usepackage{amsthm}

\usepackage{subcaption}
\usepackage{algorithm}
\usepackage{algpseudocode}

\usepackage{bigints}

\usepackage{soul}

\title{Influence of Noise on a Rotating, Softening Cantilever Beam}

\author[1]{Lautaro Cilenti\thanks{lcilenti@terpmail.umd.edu}}
\author[2]{Maria Cameron\thanks{mariakc@umd.edu}}
\author[1]{Balakumar Balachandran\thanks{balab@umd.edu}}
\affil[1]{\small{Department of Mechanical Engineering, University of Maryland, College Park, MD 20742, USA}}
\affil[2]{\small{Department of Mathematics, University of Maryland, College Park, MD 20742, USA}}

\begin{document}
\maketitle

\begin{abstract}
An experimental arrangement and a set of experiments are developed to generate empirical evidence of the effect of noise on a rotating, macro-scale cantilever structure. The experiment is a controlled representation of a rotating machinery blade. Due to the nature of the nonlinear restoring forces acting on the cantilever structure, the structure's response includes regions of multi-stability and hysteresis.  Here, a large number of trials are used to show that random perturbations can be used to create a transition between a high amplitude response and a low amplitude response of the cantilever. The observed transition behavior occurs from a high amplitude response to a low amplitude response, but not vice versa. Stochastic modeling of the system, Monte Carlo simulations, and calculations of the stochastic system's quasipotential are used to explain the nearly one-directional transition behavior.  These noise-influenced transitions can also occur in other physical systems.
\end{abstract}


{\bf Highlights}
\begin{itemize}
\item 
Construction of an experimental arrangement to study noise-influenced responses of a rotating beam structure 
\item 
First experimental demonstration of noise-induced transition from a high amplitude response to a low amplitude response of a rotating cantilever
\item 
Use of stochastic modeling, system quasipotential, and numerical studies to explain experimental observed noise-induced transition
\end{itemize}


\section{Introduction}

Systems consisting of multiple beam-like and plate-like structures, like turbomachinery, micro-electromechanical systems (MEMS), and vibration energy harvesters (VEH) can experience stable periodic mechanical oscillations that are best explained via nonlinear analysis \cite{grolet_free_2012,balachandran_response_2015,papangelo_multistability_2019}.  Proposals for new turbomachinery designs include slender, lightweight, or under-damped structures, which can give rise to high amplitude deflections of vibrating components \cite{bartels_computational_2007}; this can lead to undesired nonlinear behavior \cite{papangelo_multistability_2019}. On the other hand, this nonlinear behavior can also be desirable, for example in VEHs, in which nonlinear behavior can help enhance the frequency bandwidth over which the system response has a high energy yield \cite{jia_review_2020}. This nonlinear behavior includes regions of multiple stability; that is, the existence of a range of parameters and forces, over which these systems exhibit multiple stable nonlinear vibration responses.

It is known that random factors can induce transitions into and out of nonlinear response modes of cantilever structures. This has been demonstrated experimentally in softening cantilevers \cite{agarwal_influence_2018} and in-line arrangements of hardening cantilever beams \cite{perkins_effects_2016,alofi_noise_2022}; the cited experiments were performed using high-frequency shakers, and with a low number of trials. By contrast, the experiments in this work feature an electric motor with limited bandwidth that is more realistic for turbomachinery applications. Furthermore, these experiments are automated to create thousands of long-duration trials with low noise intensities.    

The transition behavior has also been observed in stochastic nonlinear models of cantilever structures \cite{agarwal_influence_2018,perkins_effects_2016,alofi_noise_2022,perkins_noise-influenced_2013,perkins_noise-influenced_2015,haitao_dynamics_2015,balachandran_dynamics_2022}. A common model for these structures is the harmonically forced Duffing oscillator \cite{duffing_erzwungene_1918} excited by additive white noise \cite{agarwal_influence_2018,perkins_effects_2016,balachandran_dynamics_2022}:

\begin{equation}
    \ddot{x} +\delta_c \dot{x} + \alpha {x} + \beta x^3 = F \cos(\omega t) + \sigma \eta_t,\quad x\in\mathbb{R}.
    \label{eq:Duffing}
\end{equation}

Here, $x$ is the displacement of a cantilever beam with a mass lumped at the tip from its respective equilibrium position, $\delta_c$ is a damping coefficient, and $\alpha$ and $\beta$ are scalar constants that render the oscillator monostable.  External harmonic excitations in \eqref{eq:Duffing} are described with forcing amplitude $F$, and forcing frequency $\omega$.  $\eta_t$ is the standard white noise and $\sigma$ is its scaling parameter.

Methods to quantify the probability of transitions occurring due to the influence of noise in a physical system are an active area of research \cite{alofi_noise_2022,lingala2017random,ren_local_2019,agarwal_noise-induced_2020,kikuchi_ritz_2020,chao_tao_2021,cilenti_transient_2021,zhang_koopman_2021,cilenti_most_2022}. The main challenge in accurately quantifying the stochastic nonlinear behavior is that errors compound from various sources including nonlinear model selection, system identification, noise model selection, numerical simulation errors, and  statistical errors. Some recent work has focused on nonlinear model selection, noise model selection, and system identification challenges \cite{kerschen_past_2006,breunung_noise_2022}. 

The first objective of this work is to empirically demonstrate statistically significant transition behavior. The second objective is to demonstrate that a nearly one-directional transition behavior can be observed experimentally and explained via a quasipotential analysis of the system model \cite{cilenti_most_2022}. The third objective is to show that stochastic simulations of the experimental system agree qualitatively with the experimental results. In this work, statistical errors are addressed through large number of trials. Simulation errors are addressed by comparing the results of a Monte Carlo approach; that is, Euler-Maruyama simulations of the system model \cite{higham_algorithmic_2001}, with a deterministic methodology, namely, the action plot method for non-autonomous second-order systems \cite{cilenti_most_2022}. 

The following contributions follow from this work: (i) statistically significant stochastic trials of a rotating cantilever structure transitioning between nonlinear response modes, (ii) demonstration of qualitative agreement between the experimental results and stochastic simulations of a model nonlinear system, (iii) illustration of compelling agreement between the quasipotential of the system extracted from stochastic simulations and that of the system extracted from a deterministic methodology, and (iv) collection of empirical evidence correlating the observed transition behavior and the quasipotential of the system at different excitation frequencies.   

The paper is organized as follows: The experimental arrangement is presented in Section \ref{Section_OneCantileverExperiment}. The frequency response of the cantilever beam and the system identification approach to extract a comparable model is discussed in Sections \ref{Section_OneCantileverFrequencyResponse} and \ref{Section_OneCantileverSystemIdentification}, respectively. Experimental results are presented in Section \ref{Section_OneCantileverExperimentalResults}, and later, qualitatively compared to simulation results in Section \ref{Section_OneCantileverSimulationResults}. An analysis demonstrating agreement between the quasipotential found stochastically via numerical trials and the quasipotential found via a deterministic methodology for this system is discussed in Section \ref{Section_QuasipotentialAnalysis}.  Concluding remarks are presented in Section \ref{Section_OneCantileverDiscussion}.

\section{Experimental Arrangement} \label{Section_OneCantileverExperiment}

\begin{figure}[htbp]
\centerline{\includegraphics[width=0.5\textwidth]{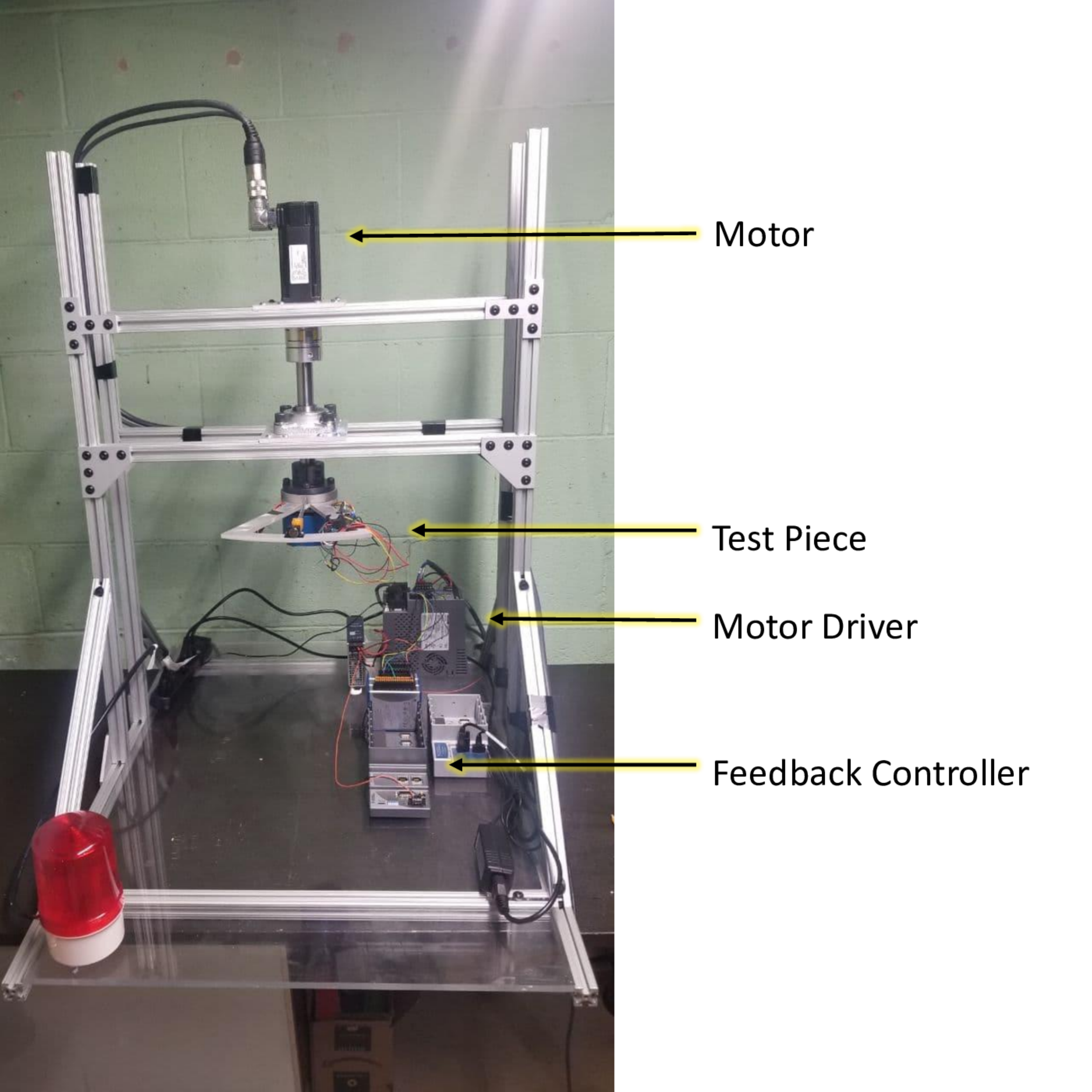}}
\caption{Experimental prototype with one cantilever beam.}
\label{Exp_Prototype1}
\end{figure}

The experimental setup consists of an excitation system, a data acquisition system, a test piece, and mechanical supports.
An image of the prototype is shown in Fig. \ref {Exp_Prototype1}. The excitation system consists of a feedback controller, a motor driver, and a motor. The control is provided via an FPGA module in an NI-cRIO9074 hardware. The FPGA module is used to update a voltage output at 80 kHz in order to generate harmonic excitations and additive random perturbations. 
\noindent
This module is also used to collect data at $5$kHz from a strain gauge and store it locally. 
A Kollmorgen AKM motor is used to deliver torque via a shaft; this generates accelerations of the base that vibrate the test piece. By nature and design, the motor is not a high-frequency excitation system. The motor acts as a low pass filter, and in the motor response, the high-frequency components of the reference signal are cut off. In comparison to an electrodynamic shaker designed for high-frequency excitations, this cutoff frequency is lower for the motor. The low pass filtering effect of the rotating motor at the excitation frequencies, which is similar to the real applied use cases of the considered rotating cantilever arrays, can be difficult to model. 

\begin{figure}[htbp]
\centerline{\includegraphics[width=0.5\textwidth]{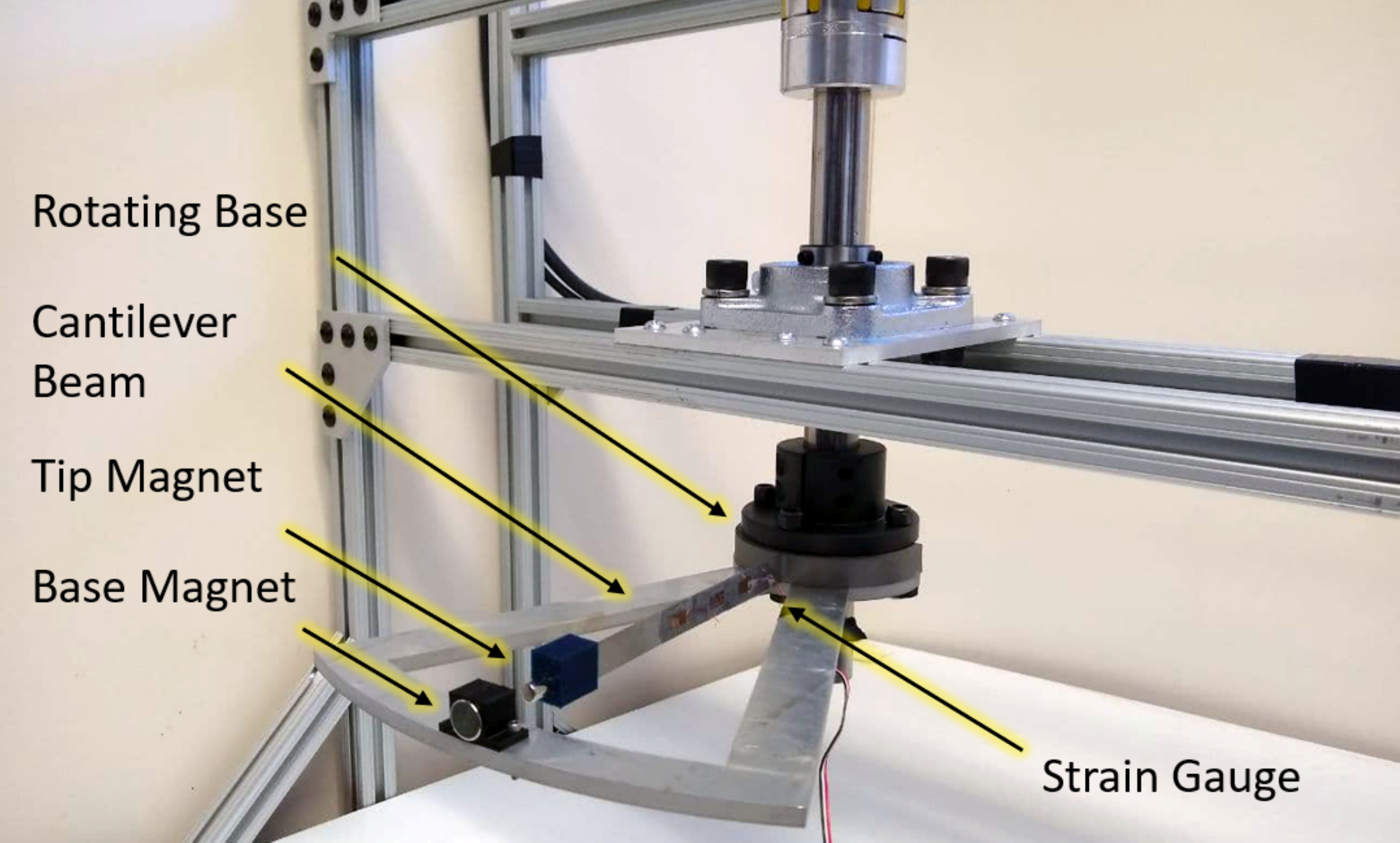}}
\caption{Test piece prototype with one cantilever beam.}
\label{DCL_CantileverBeamOnRotatingBase}
\end{figure}

The test piece is the cantilever steel structure shown in Fig. \ref {DCL_CantileverBeamOnRotatingBase}. The cantilever is 15.24 cm (6.00 inches) in length. A permanent magnet is attached to the tip of the cantilever beam and held in place by a 3D-printed component. A strain gauge, which is located 3.81 cm (1.50 inches) from the base, is used to estimate the displacement of the cantilever tip. A larger, more rigid structure under the cantilever beam holds a second permanent magnet. This structure moves as a rigid body with the shaft. The interaction between the magnets creates a nonlinear restoring force, primarily affecting the position of the tip of the cantilever beam. 

\section{Frequency Response} \label{Section_OneCantileverFrequencyResponse}

\begin{figure}[htbp]
\centerline{\includegraphics[width=0.5\textwidth]{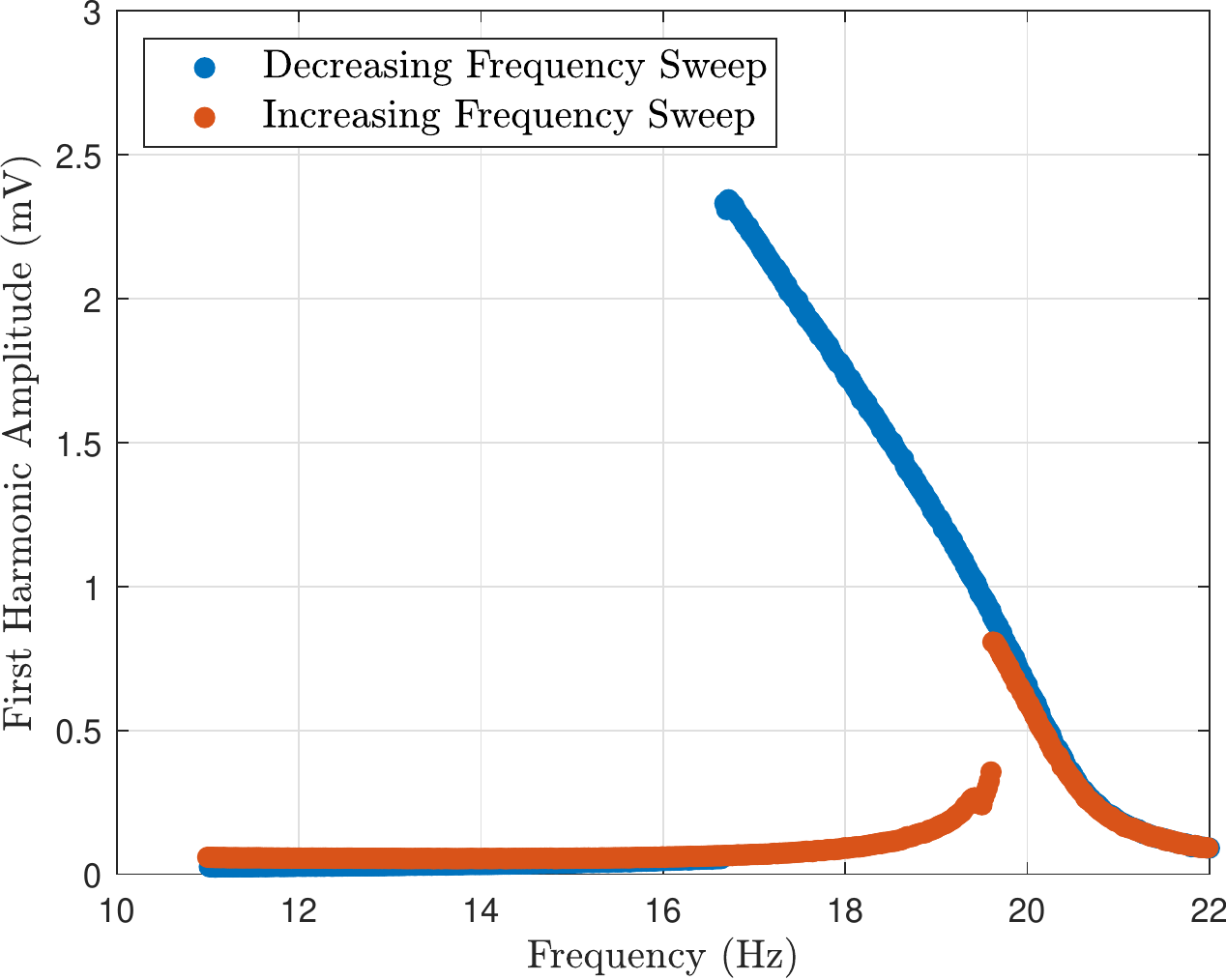}}
\caption{First harmonic amplitude of the response, $a$, as a function of frequency, $f$.}
\label{OneCantilever_FrequencyResponse}
\end{figure}

One way to capture the nonlinear behavior of the system is by performing a frequency sweep. The frequency sweep consists of quasi-statically changing the excitation frequency; that is, changing and sustaining each excitation frequency for a sufficient duration so that the steady state oscillations of the system can be recorded. The obtained frequency response of the system is shown in Fig. \ref {OneCantilever_FrequencyResponse}.

Notice from Fig. \ref {OneCantilever_FrequencyResponse} that between $f \approx 16.60$ Hz and $f \approx 19.60$ Hz, that there exist two stable steady state responses of the system at each excitation frequency. The high amplitude response occurs in the region shaded in blue and the low amplitude response occurs in the region shaded in red. A noise induced transition between high amplitude vibrations and low amplitude vibrations is simply the system response state switching from the red region to the blue region or vice versa due to random external perturbations.

\section{System Identification} \label{Section_OneCantileverSystemIdentification}

In this section, a system identification approach to curve fit experimental data to the stable frequency response of Duffing oscillators is presented. 
This approach expands on the system identification method described by Agarwal et al. \cite{agarwal_influence_2018}.

Parametric system identification can be performed by fitting the  experimental data in Fig. \ref {OneCantilever_FrequencyResponse} to the nondimensionalized frequency response of a forced Duffing oscillator \cite{nayfeh_nonlinear_2008}:
\begin{equation}
    \left[ \frac{\eta^2}{4}  + 
    \left(\Omega - 1 -  \frac{3}{8} \beta A^2\right)^2 \right] 
    = \frac{\hat{F}_0^2}{4  A^2}
    \label{NDAnalyticalSpectrum}
\end{equation}
Here $\eta$, $\Omega$, $\beta$, and $\hat{F_o}$ are nondimensionalized damping, frequency, cubic coefficient, and excitation amplitude, respectively. The parameter $A$ is the nondimensional amplitude defined as $A = a(\Omega)/a_p$, where $a(\Omega)$ is the amplitude of the system response at a specific $\Omega$, and $a_p$ is the supremum of $a(\Omega)$. One can extract dependence of the nondimensional frequency $\Omega$ on $A$ from \eqref{NDAnalyticalSpectrum}; that is:

\begin{equation}
\label{eq:Omega}
    \Omega 
    = 1 +  \frac{3}{8} \beta A^2 \pm \sqrt{\frac{\hat{F}_0^2}{4  A^2} - \frac{\eta^2}{4}} 
\end{equation}
The supremum,  $a(\Omega) = a_p$, corresponding to $A=1$,  occurs at the jump down frequency. 
For $A>1$ solutions of the analytical curve are complex numbers. The transition to complex numbers occur when $\sqrt{\frac{\hat{F}_0^2}{4  A^2} - \frac{\eta^2}{4}} = 0$, which for $A = 1$ occurs when $\hat{F}_0 = \eta$. Hence, it is assumed $\eta$  is equal to $\hat{F}_0$ to reduce the dimensionality of the regression problem and constrain the regression problem to parameters that correspond to a real response up to $A=1$.
The remaining parameters, $\beta$ and $\hat{F}_o$, are found by fitting the experimental nondimensional  frequencies $\Omega_i$ and the corresponding measured nondimensional first harmonic amplitudes $A_i$ to ansatz \eqref{eq:Omega}. The cost function for this optimization problem is chosen to be
\begin{equation}
    \begin{split}
       & C(\beta,\hat{F}_0)  = \sqrt{\sum_{i=1}^{N}{\left(1 +  \frac{3\beta A_i^2}{8}  + s_i \sqrt{\frac{\hat{F}_0^2}{4  A_i^2} - \frac{\hat{F}_0^2}{4}} - \Omega_i\right)^2}}\\ 
        &s_i  = \begin{cases} 1, & \textrm{sample $i$ fits high amplitude branch} \\ 
        -1, & \textrm{sample $i$ fits low amplitude branch} \\ 
        \end{cases}
    \end{split}
    \label{eq_OneCantilever_AnalyticalSysIDCost}
\end{equation}
 
Solving \eqref{eq_OneCantilever_AnalyticalSysIDCost} is a nonlinear regression problem with respect to $\beta$ and $\hat{F}_0$. This is a low-dimensional problem that is convenient to solve with common nonlinear optimization packages. The found parameter values are $\beta = -0.48$ and  $\hat{F}_0 = \eta = 0.0104$.

\begin{figure}[htbp]
\centerline{\includegraphics[width=0.5\textwidth]{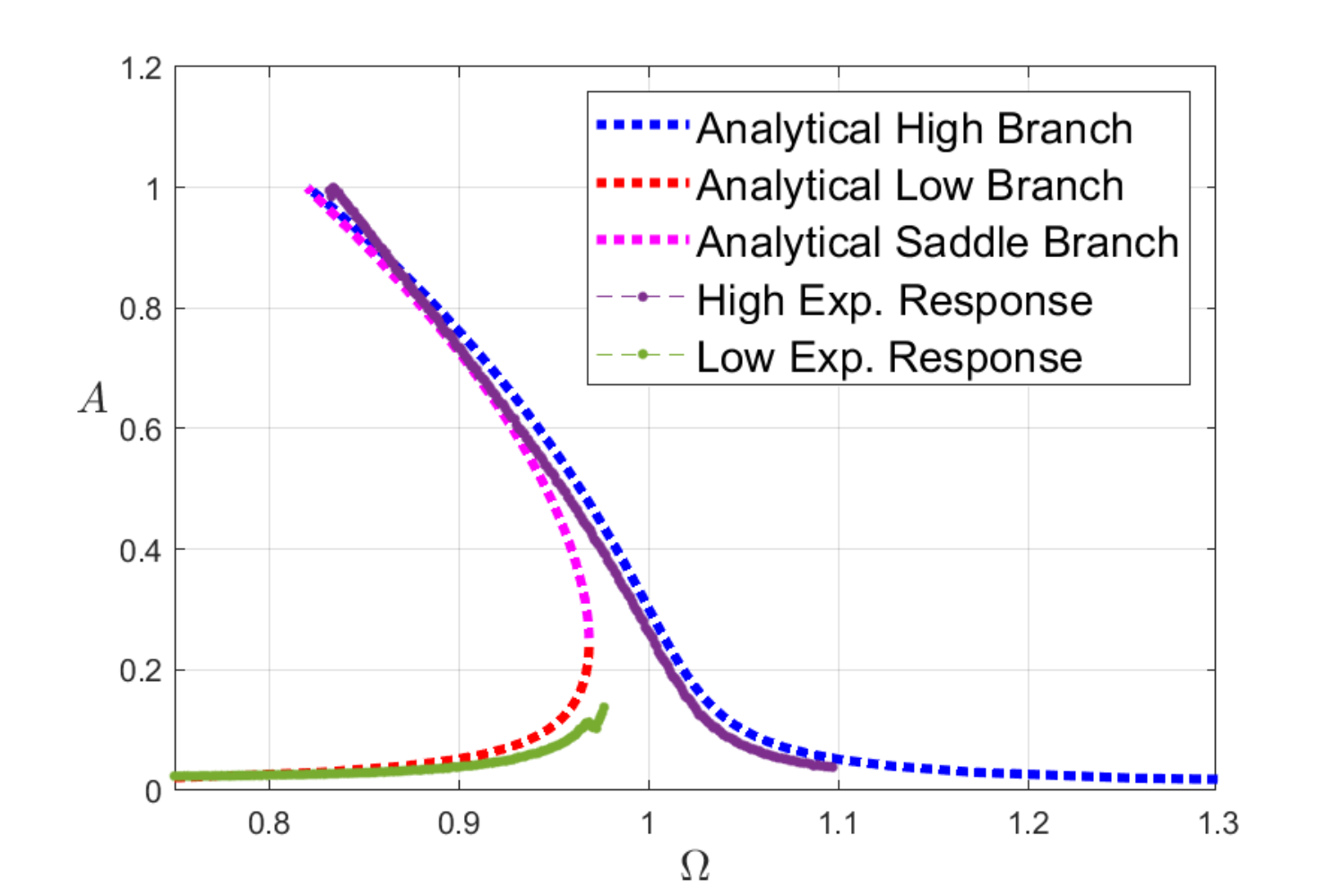}}
\caption{The experimental frequency response in nondimensionalized form, compared to the analytical approximation to the frequency response of the Duffing oscillator with parameters that minimize \eqref{eq_OneCantilever_AnalyticalSysIDCost}. Here, $\beta = -0.48$,  $\hat{F}_0 = \eta = 0.0104$.}
\label{OneCantilever_AnalyticalSysID}
\centerline{\includegraphics[width=0.5\textwidth]{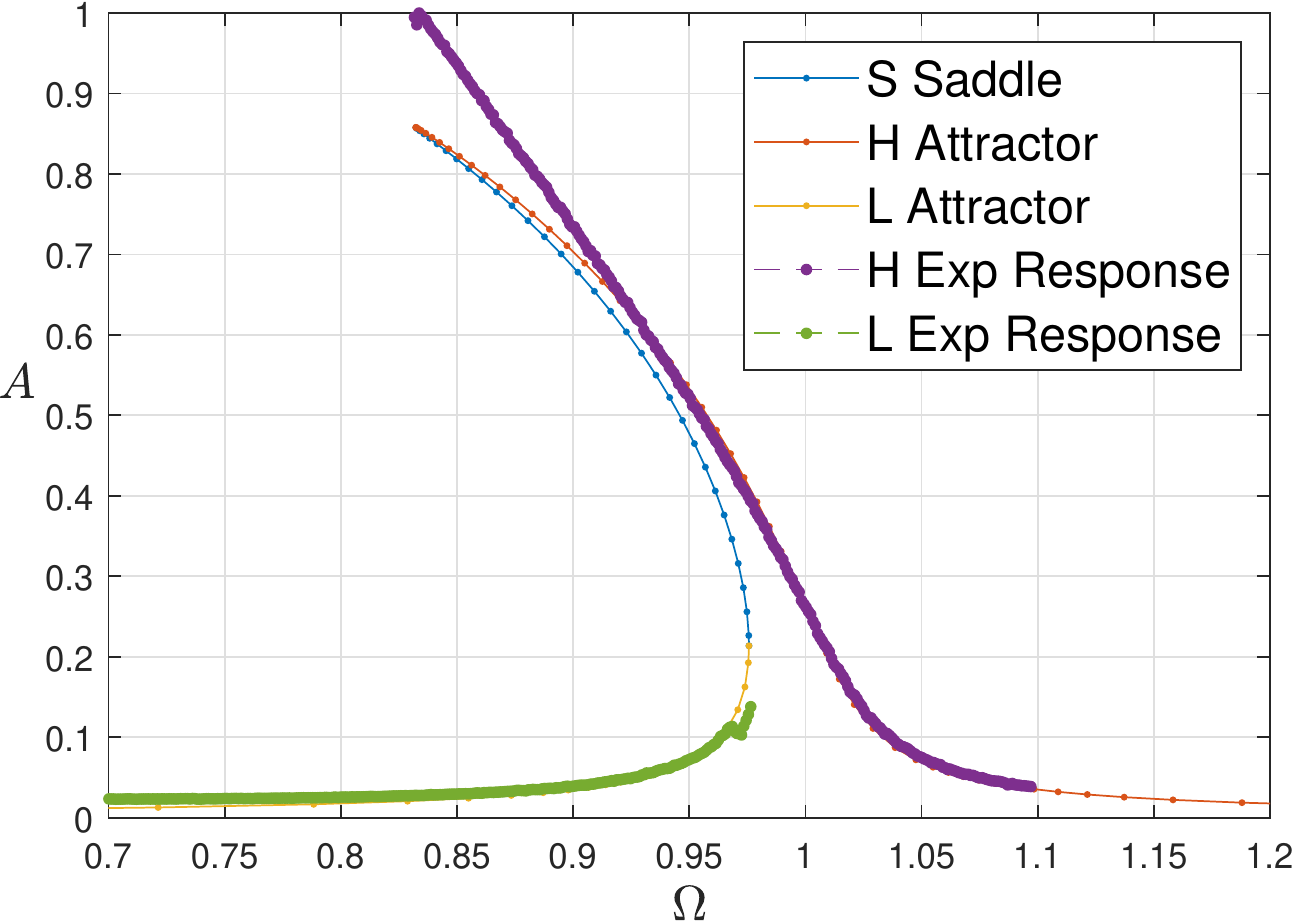}}
\caption{The experimental frequency response in nondimensionalized form compared to the frequency response of the Duffing oscillator computed via numerical continuation. Here, $\beta = -0.48$,  $\hat{F}_0 = 0.007$, and  $\eta = 0.009$. }
\label{OneCantilever_COCOSysID}
\end{figure}

The experimental frequency response curve from Fig. \ref{OneCantilever_FrequencyResponse} put in the nondimensional form and the analytical response curve for the found parameter values $\beta = -0.48$ and  $\hat{F}_0 = \eta = 0.0104$ are presented together in Fig. \ref{OneCantilever_AnalyticalSysID}.  It is noted that the analytical curve includes a saddle branch that corresponds to unstable solutions that are not observed in the experiments.


It is important to note that the analytical approximation for the frequency response \eqref{NDAnalyticalSpectrum} 
derived in \cite{nayfeh_nonlinear_2008} is based on the assumption of weak nonlinearities, but the nonlinear system \eqref{eq:Duffing} with cubic coefficient $\beta = -0.48$ is not weakly nonlinear.  
The shooting method and numerical continuation of periodic solutions are used to generate the frequency response of the system without the weakly nonlinear assumptions, and the parameters are adjusted to refit the numerical response. The numerical continuation package COCO \cite{dankowicz_coco_2019} is used for this step. Since numerical continuation is computationally expensive,  it is only used as a minor parameter correction step and not as the sole system identification approach. 

The frequency response obtained by using a numerical continuation is shown alongside the experimental data in Fig. \ref {OneCantilever_COCOSysID}. The corrected parameters are $\beta = -0.48$,  $\hat{F}_0 = 0.007$, and  $\eta = 0.009$.
In Fig. \ref {OneCantilever_COCOSysID}, the experimental data diverges from the best fit model response at high amplitudes. This happens with both the initial set of parameters minimizing \eqref{eq_OneCantilever_AnalyticalSysIDCost} and after the additional minimization with numerical continuation. The representation and fit in Fig. \ref {OneCantilever_COCOSysID} is preferred because there is good agreement of the bifurcation points, and because the response of the numerical continuation approximation is more accurate than the response obtained by using the analytical approximation.  Additional nonlinear terms and/or modeling of the cantilever structure with a higher dimensional system (i.e., more degrees of freedom) can help explain the system high amplitude behavior.

\section{Experimental Results} \label{Section_OneCantileverExperimentalResults}

\begin{figure}[htbp]
\centerline{\includegraphics[width=0.5\textwidth]{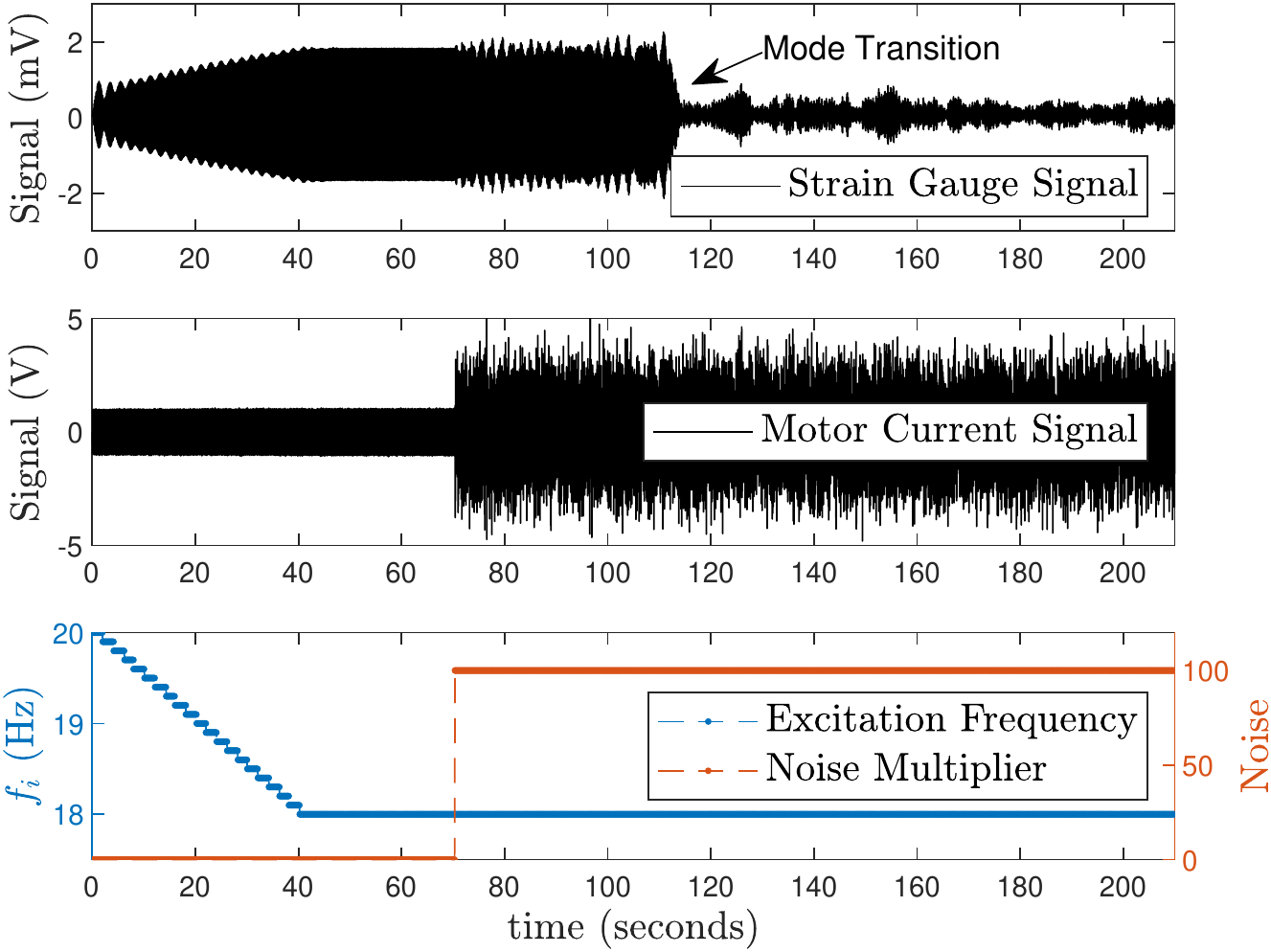}}
\caption{An example noise trial in the experimenst. Each trial was started with a gradual decrease in the excitation frequency, $f_i$, which leads the strain response to a high amplitude stable mode. At $t \approx70$ seconds, noise is added to the control signal. After some time, the random perturbations induced a transition from the high amplitude strain response to a low amplitude strain response.}
\label{OneCantilever_TrialTimeSeries}
\end{figure}

\begin{figure}[htbp]
\centerline{\includegraphics[width=0.5\textwidth]{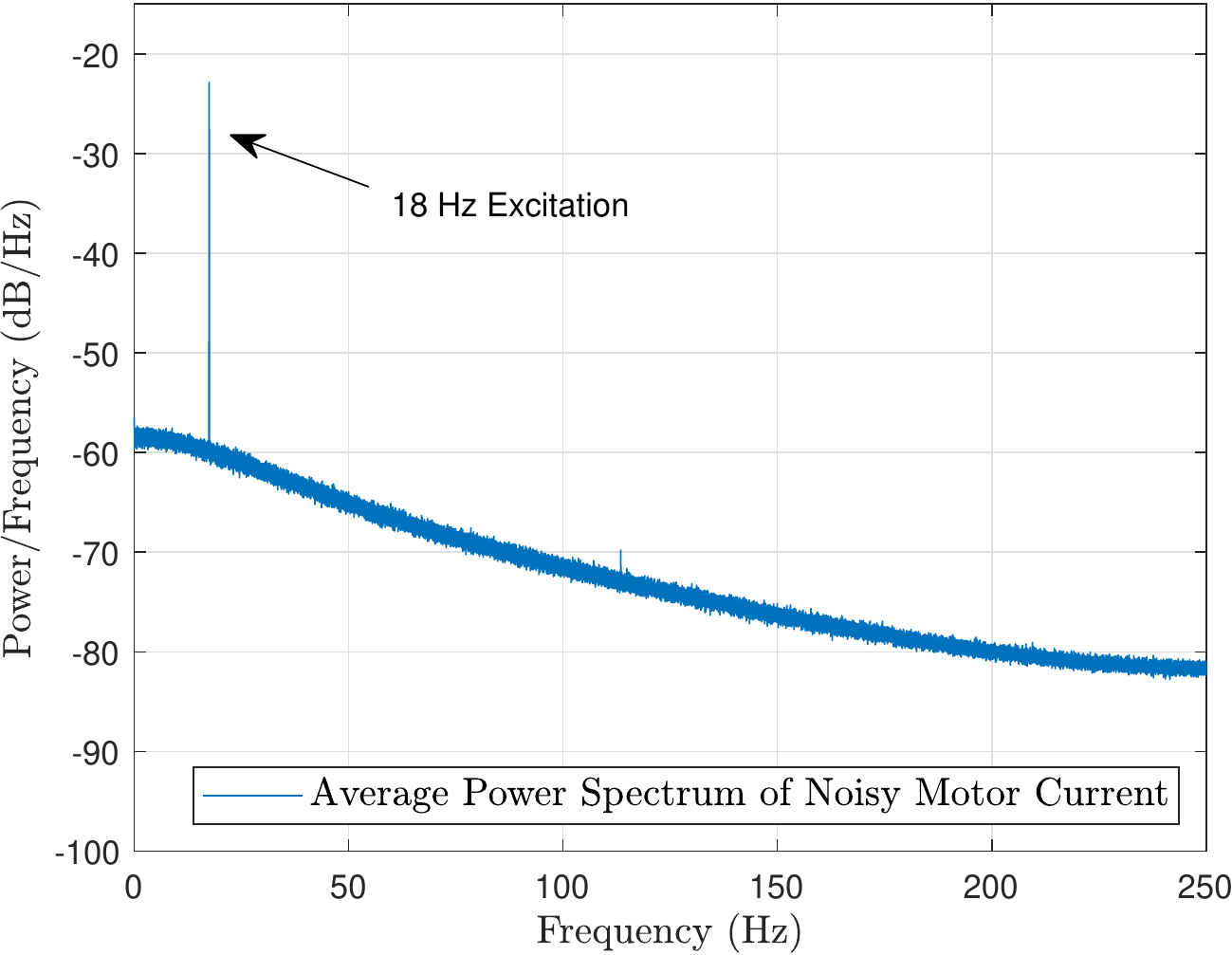}}
\caption{The power spectrum of the noisy control signal averaged over $500$ trials for $\sigma_E = 100$ . Notice a sharp peak at $18.00$ Hz corresponding to the deterministic cosine excitation and attenuation of the signal at higher frequencies that occurs due to the low pass behavior of the motor.}
\label{OneCantilever_NoiseAveragePowerSpectrum}
\end{figure}

The experiment is used to create empirical evidence of transitions occurring from the high amplitude mode to the low amplitude mode due to the effect of noise. A noise trial in the experiment takes the form shown in Fig. \ref {OneCantilever_TrialTimeSeries}, which consists of plots of the strain, the motor current, the excitation frequency parameter $f_i$, and the noise multiplier parameter, $\sigma_E$. First, note that in the first 70.00 seconds of a trial, $\sigma_E = 0$, meaning there is no additive noise being intentionally added to the system. In the first 40.00 seconds, the excitation frequency $f_i$ is slowly decreased from 20.00 Hz to 18.00 Hz. It is noted that as the excitation frequency decreases, the strain increases; the system follows the blue curve of Fig. \ref {OneCantilever_FrequencyResponse} to higher amplitude responses. At the forty seconds mark, the frequency of excitation is fixed to the constant value of $18.00$ Hz. Between $40.00 \leq t \leq 70.00$, no parameters are changed, and the strain reaches a steady state oscillation. For $t > 70.00$, $\sigma_E$ is set to $100$. 
Notice that the motor current at $\sigma_E$ transitions from smooth sinusoidal oscillations to noisy oscillations at $t = 70.00 $ seconds. The average power spectrum of the noisy control signal is shown in Fig. \ref {OneCantilever_NoiseAveragePowerSpectrum}. There is a peak of the control signal at 18.00 Hz corresponding to the deterministic cosine excitation. The roll off is a low pass filter effect on the noise.

Despite the seemingly high amount of noise, the strain response does not immediately transition from the high amplitude mode to the low amplitude mode. A transition is depicted in Fig. \ref {OneCantilever_TrialTimeSeries} at $t \approx 110.00$ seconds, which is after approximately $720.00$ periods of excitation since the noise is introduced. The trial always ends at $t \approx 225.00$ seconds, regardless of whether a transition has been observed. The full duration of each trial is run to capture the possibility for the system to transition back to the high amplitude mode; a transition from the low amplitude response to the high amplitude response was never observed in any of the trials. 
Only a single trial is depicted in Fig. \ref {OneCantilever_TrialTimeSeries}. The controller is programmed to repeat these trials automatically 500 times for one set of excitation frequency and noise multiplier parameters. A set of 500 trials takes between $35$ to $40$ hours to complete.  Due to the nature of the random perturbations in the control signal, the transition between the high amplitude mode and the low amplitude mode occurs at different times in each trial. 

\begin{figure}[htbp]
\centerline{\includegraphics[width=0.5\textwidth]{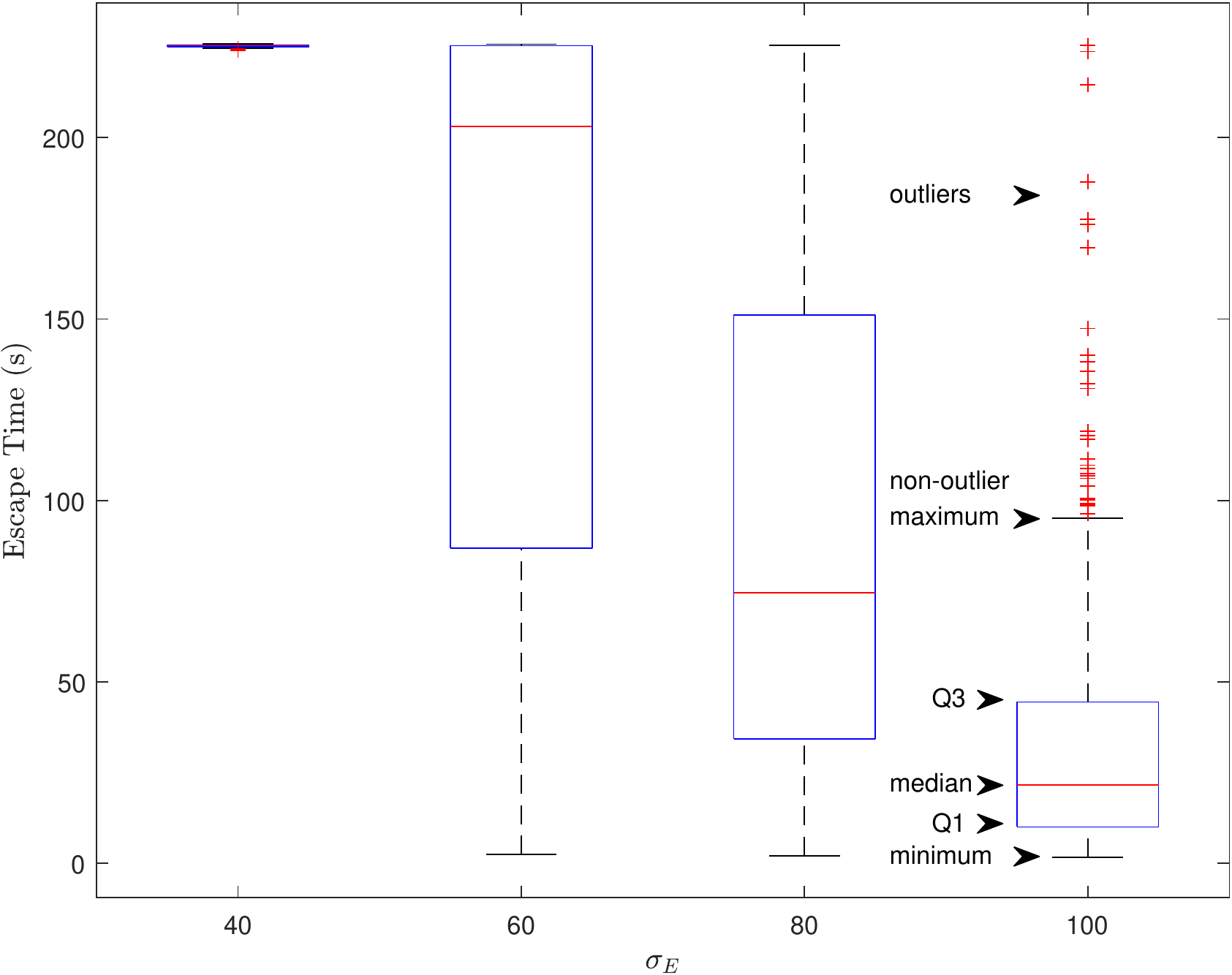}}
\caption{Box plots of the noise trials. Each box plot consists of 500 experimental trials for a different value of the noise multiplier. Here, for the deterministic cosine excitation, the excitation frequency is $18.00$ Hz.}
\label{OneCantilever_EscapeTimes18Hz}
\end{figure}

Box plots of the transition time of 500 trials of different noise intensities are shown in Fig. \ref {OneCantilever_EscapeTimes18Hz}. One can observe roughly an exponential relationship between the median transition time and a linear change in the noise intensity. 

\begin{figure}[htbp]
\centerline{\includegraphics[width=0.5\textwidth]{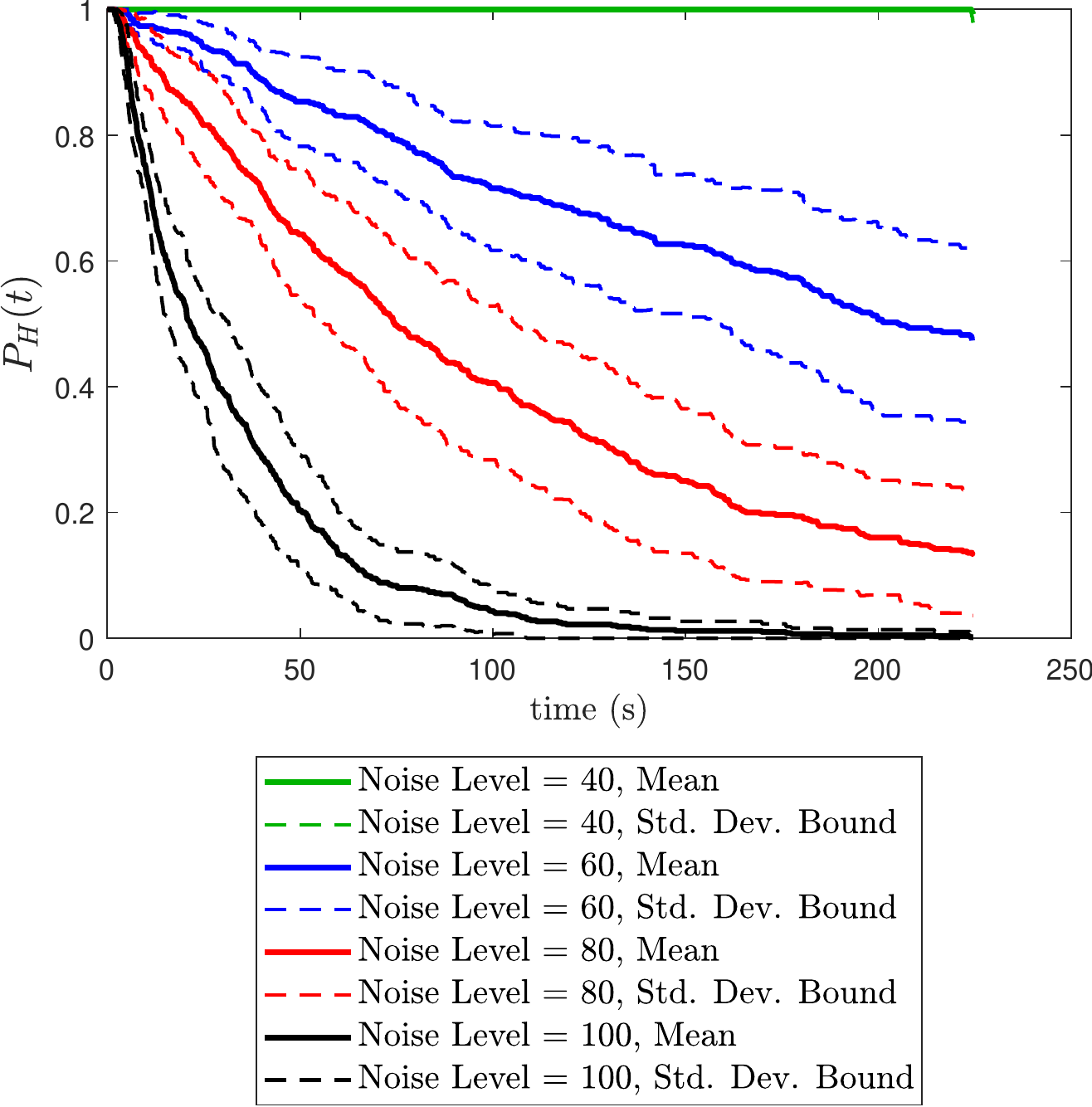}}
\caption{Estimate of the probability of remaining in the high amplitude attractor as a function of time and noise intensity, $\sigma_E$. Here, the excitation frequency is fixed at $f_i = 18.00 $ Hz. These curves are created by using \eqref{OneCantilever_ProbabilityEstimate} on the experimental escape times.}
\label{OneCantilever_ExperimentalProbabilityEstimate18Hz}
\end{figure}

An alternative interpretation of this data can be expressed as the probability of the system remaining in the high amplitude response for different noise intensities and durations. By using the escape times, this probability can be estimated as follows:
\begin{equation}
\begin{split}
        P_H(t) & = \frac{\sum_{i = 1}^{n} \delta_e(t,\tau_i)}{n} \\ 
        \delta_e(t,\tau_i) & = 
        \begin{cases}
        1, \quad t< \tau_i\\
        0, \quad t\geq \tau_i
        \end{cases}
\end{split}
\label{OneCantilever_ProbabilityEstimate}
\end{equation}
Here, $\tau_i$ is the escape time of trial $i$, and $\delta_e$ is a function that gives a trial a value of $1$ at times before the response has transitioned out of the high amplitude response, and zero otherwise. The variable $n$ is the bin size of trials in the estimate. The probability estimates for different noise intensities at the excitation frequency of $18$ Hz are shown in Fig. \ref {OneCantilever_ExperimentalProbabilityEstimate18Hz}. In Fig. \ref {OneCantilever_ExperimentalProbabilityEstimate18Hz}, $n=50$, and trials are randomly selected from the set of $500$ trials for the calculation. A total of $10$ estimates are determined  for each noise intensity, and the resulting $10$ estimates are averaged and graphed alongside with the standard deviation as an indicator of the error of the estimate. In Fig. \ref {OneCantilever_ExperimentalProbabilityEstimate18Hz}, on the solid green curve corresponding to $\sigma_E = 40$, the system has a high probability of remaining in the high amplitude response. On the black curve corresponding to $\sigma_E = 100$, the system has the lowest probability of remaining in the high amplitude response after a short amount of time. The other curves follow the same trend with varying noise intensity.  

\begin{figure}[htbp]
\centerline{\includegraphics[width=0.5\textwidth]{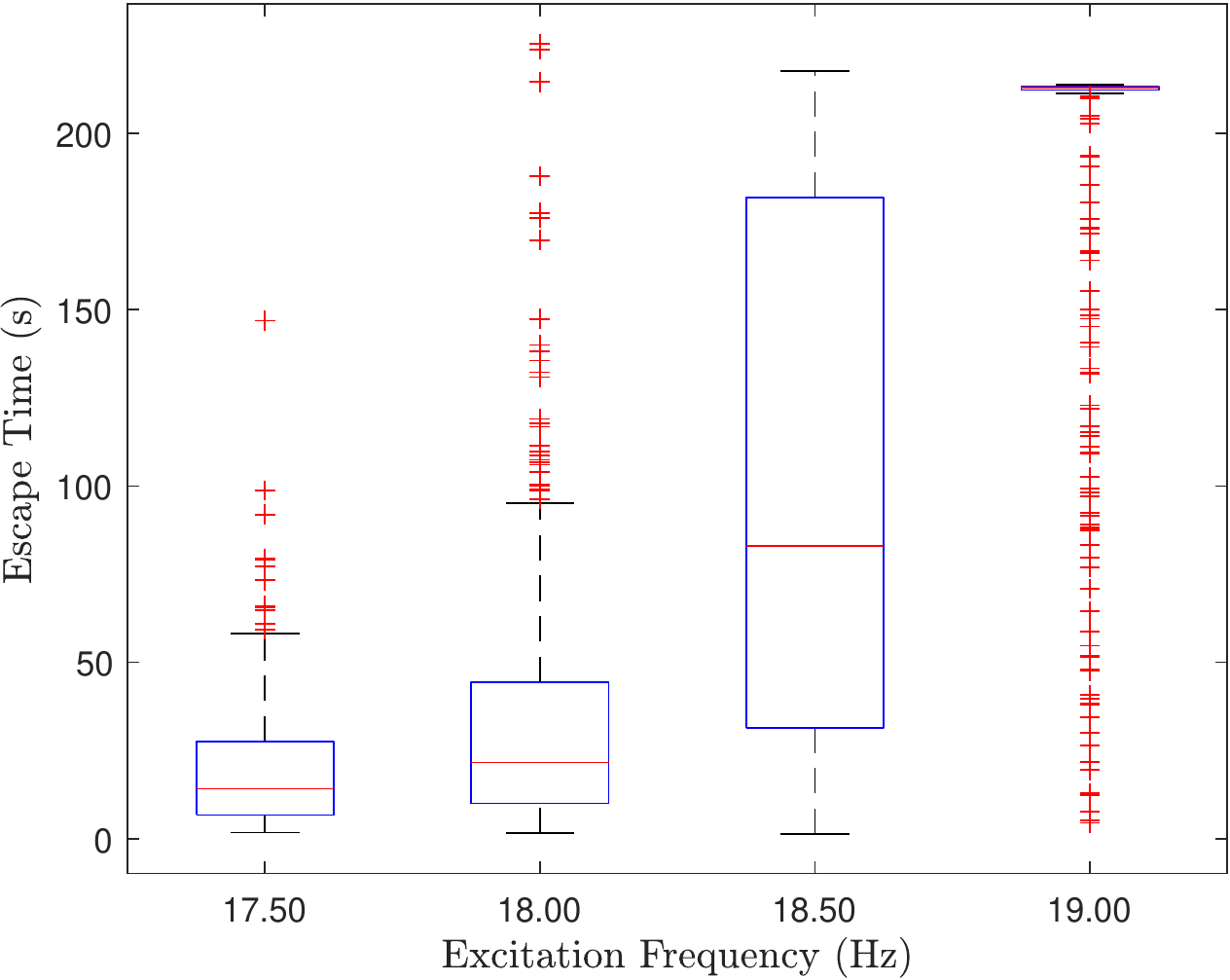}}
\caption{Box plots of the noise trials. Each box plot consists of 500 experimental trials for a different value of the excitation frequency, $f_i$. Here $\sigma_E = 100$. }
\label{OneCantilever_EscapeTimesNoise100}
\end{figure}

\begin{figure}[htbp]
\centerline{\includegraphics[width=0.5\textwidth]{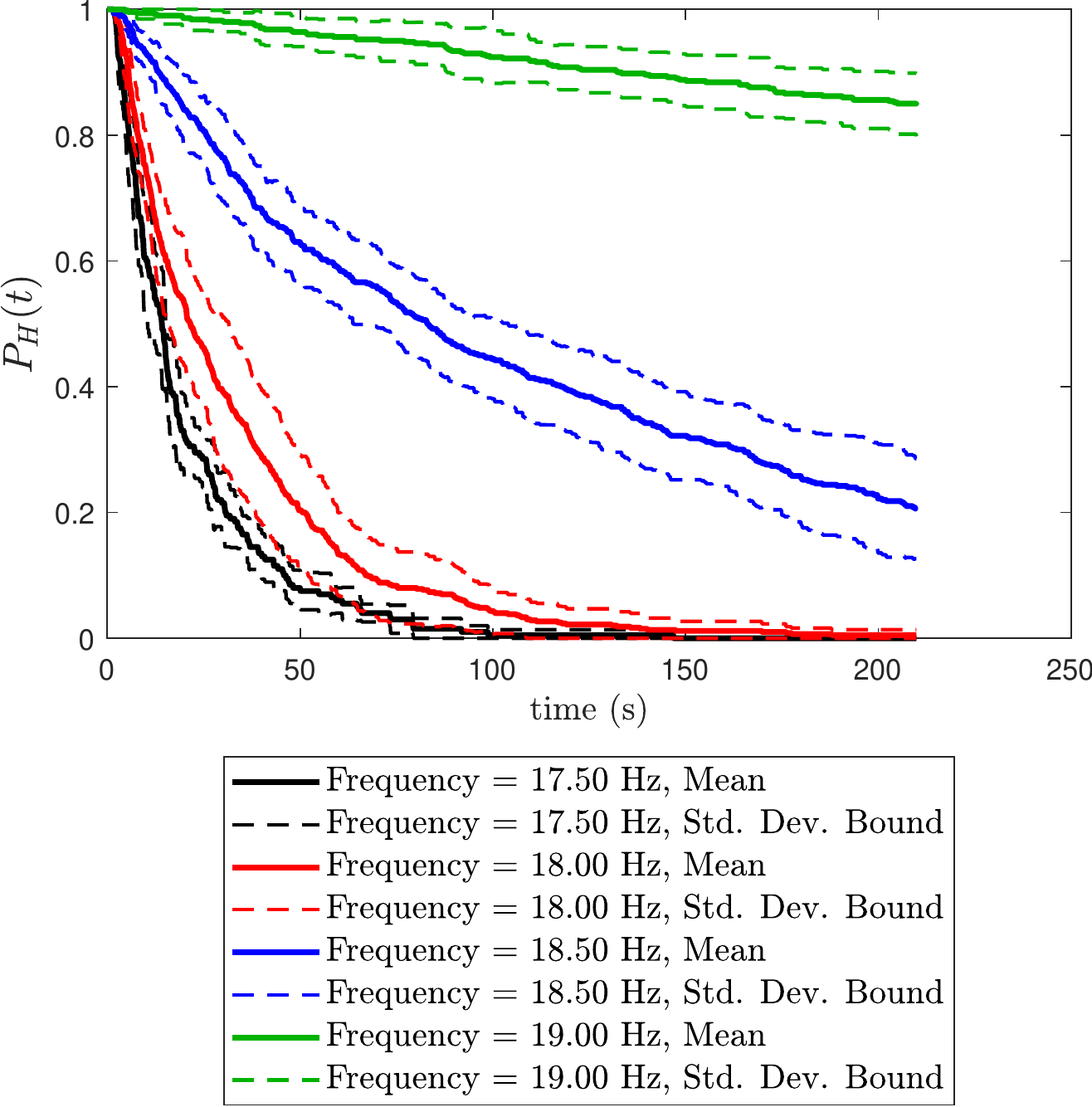}}
\caption{Estimate of the probability of remaining in the high amplitude attractor as a function of time and excitation frequency. Here, the noise intensity is fixed at $\sigma_E = 100$. These curves are created by using \eqref{OneCantilever_ProbabilityEstimate} on the experimental escape times.}
\label{OneCantilever_ExperimentalProbabilityEstimateN100}
\end{figure}

The dependence of the escape times and the probability estimates on the excitation frequency is also studied. $500$ trials are performed for each of four excitation frequencies in the multi-stability region of the frequency response. The noise intensity is kept constant at $\sigma_E = 100$. Box plots of the escape times from the high amplitude response as a function of noise intensity are shown in Fig. \ref {OneCantilever_EscapeTimesNoise100}. One can observe in Fig. \ref {OneCantilever_EscapeTimesNoise100} an exponential relationship between the median escape time with linear changes in excitation frequency. In particular, for the excitation frequencies of $17.50$, $18.00$, and $18.50$ Hz, the majority of trials escape the high amplitude response within the $5000$ period maximum time limit. On the other hand, for the excitation frequency of $19.0$ Hz, the majority of trials do not escape within the first $5000$ periods. These results support observations in related work \cite{alofi_noise_2022}, which indicate that transitions from the high amplitude response to the low amplitude response are more probable near the jump down point of the frequency response, and less probable near the jump up point of the frequency response. 
 The probability estimates for the system to remain in the high amplitude attractor at excitation frequencies of 17.50, 18.00, 18.5, and 19.00 Hz are displayed in  Fig. \ref {OneCantilever_ExperimentalProbabilityEstimateN100}. A significant dependence of the probability curves on the excitation frequency is evident.
%

\section{Simulation Results}
\label{Section_OneCantileverSimulationResults}

\begin{figure}[htbp]
\centerline{\includegraphics[width=0.5\columnwidth]{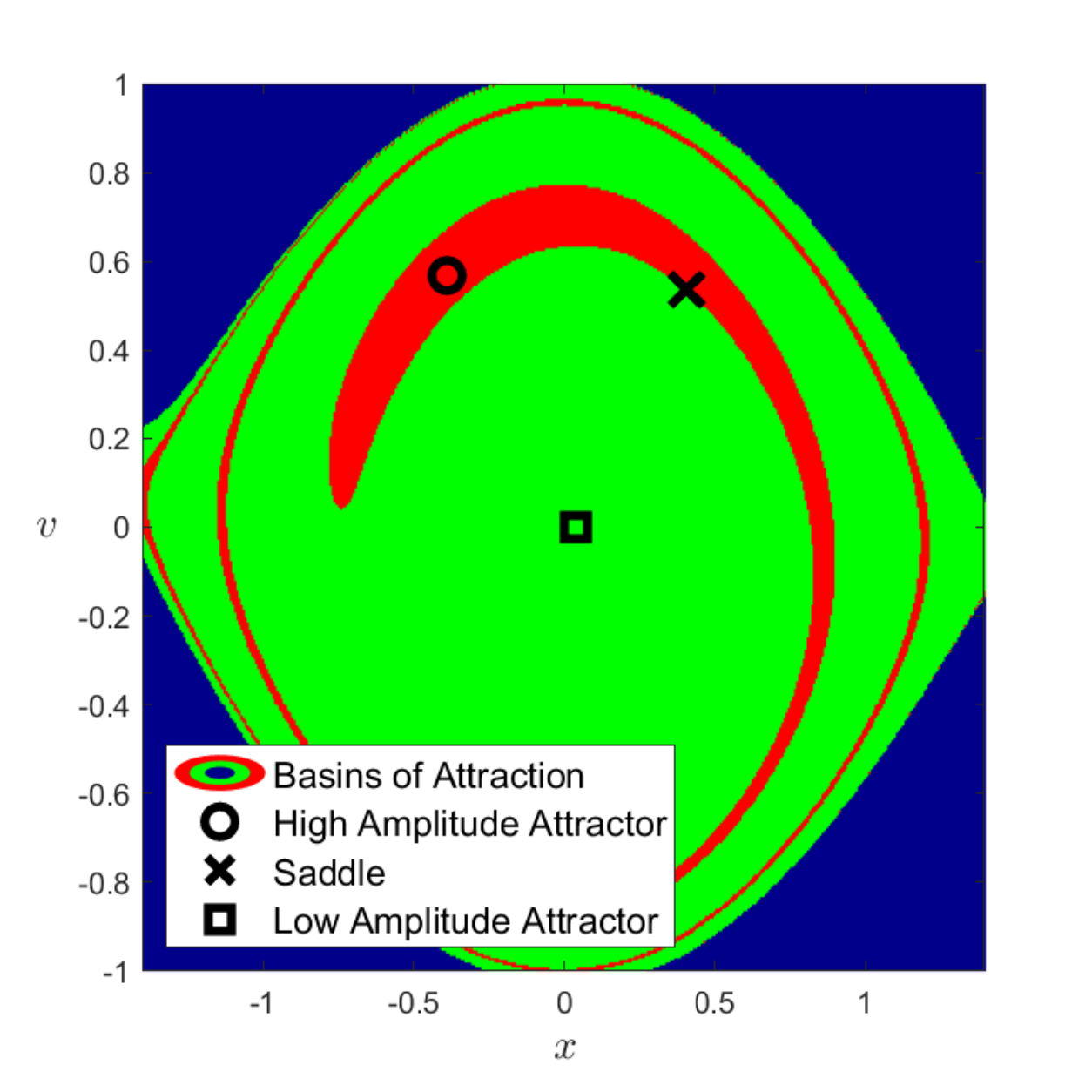}}
\caption{Basins of attraction of the system model at $\Omega = 0.9$ or equivalently $f_i = 18.00 $Hz. The basin of the high amplitude attractor and the low amplitude attractor are shown in red and green, respectively. An unstable basin in which the response blows up to infinity is shown in blue. }
\label{OneCantilever_18HzBasins}
\end{figure}

The system identification approach in Section \ref{Section_OneCantileverSystemIdentification} produced the nondimensional ordinary differential equation (ODE)
\begin{equation}
     x''+ 0.009 x' + x - 0.48 x^3 = 0.007 \text{cos}(\Omega \tau).
     \label{eq_OneCantileverODE}
\end{equation}
Including an additive white noise excitation, the corresponding stochastic differential equation (SDE) is 
\begin{equation}
     x''+ 0.009 x' + x - 0.48 x^3 = 0.007 \text{cos}(\Omega \tau) +\sigma \eta_\tau.
     \label{eq_OneCantileverSDE}
\end{equation}
Here, $\sigma$ is the noise intensity multiplier and $\eta_\tau$ is the additive white noise. The excitation frequency $f_i = 18$ Hz is equivalent to $\Omega = 0.9$ in nondimensional form. In Fig. \ref {OneCantilever_18HzBasins}, the Poincar{\'e} sections of the basins of attraction of \eqref{eq_OneCantileverODE} corresponding to  phase zero; that is, $\Omega\tau \mod 2\pi = 0$, for $\Omega = 0.9$ are shown.  For brevity, these Poincar{\'e} sections of basins of attractions are just referred to as basins. The high amplitude and low amplitude attractors have the basins shown in red and green, respectively.  The blue region corresponds to initial conditions that are unstable and grow to infinity. Unstable vibrations that grow to infinity do not occur in the experiment, so it is assumed the model is valid and to be used within the domains of the green and red regions. For low $\sigma$ and initial conditions of ODE \eqref{eq_OneCantileverODE} near one of the two attractors, the system with high probability remains in the red or green regions and does not enter the unrealistic unstable domain. 
Transitions in the system model from the high amplitude attractor to the low amplitude attractor are responses of the system in which the system starts near the high amplitude attractor (the circle of Fig. \ref {OneCantilever_18HzBasins}), moves towards the green basin of attraction via the work performed by random perturbations, and then moves along trajectories of \eqref{eq_OneCantileverODE} to the square of Fig. \ref {OneCantilever_18HzBasins}. Note that the high amplitude basin is much smaller than the low amplitude basin in size.

\begin{figure}[htbp]
\centerline{\includegraphics[width=0.5\textwidth]{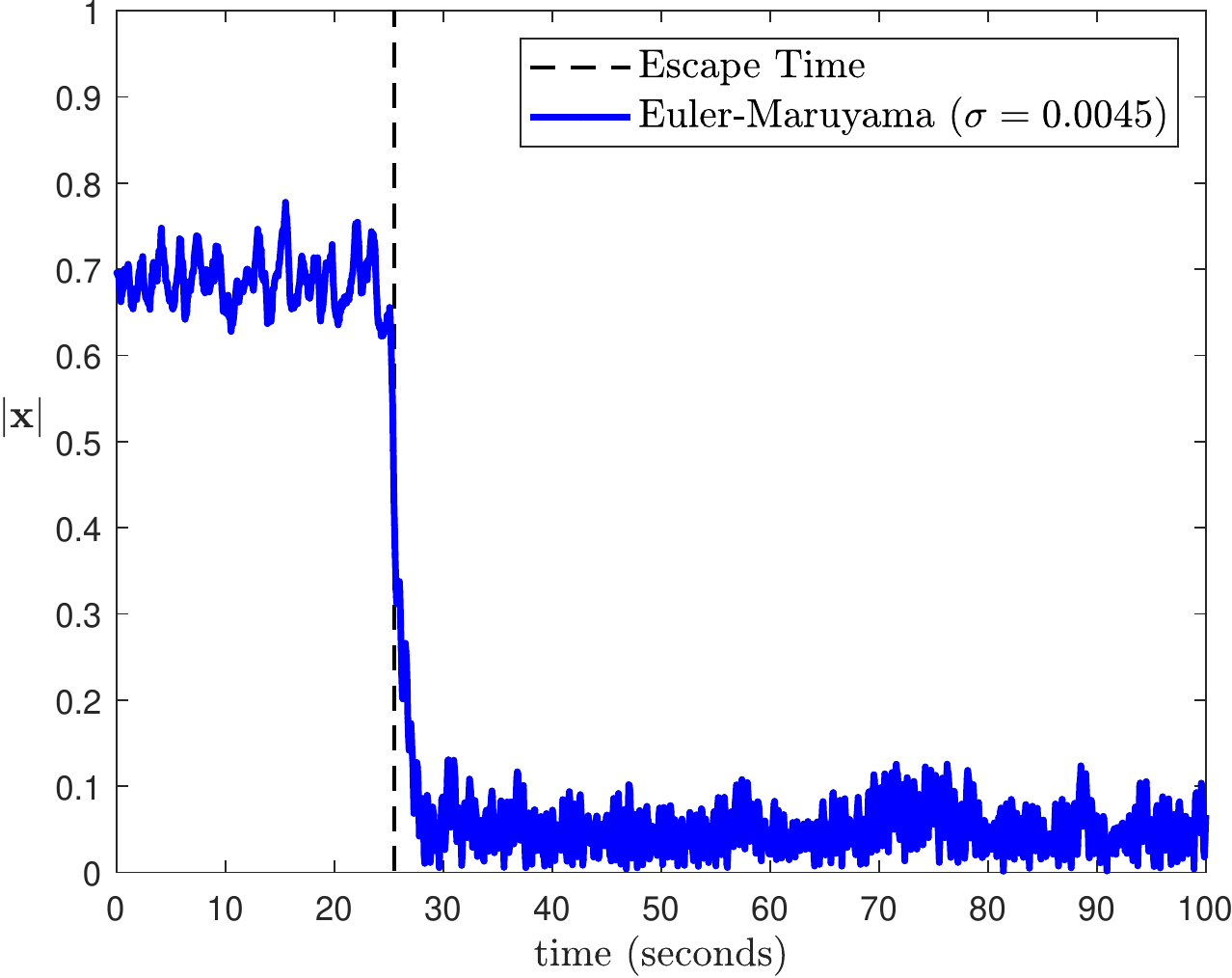}}
\caption{The Euclidean norm of the Poincar\'e section of the stochastic response of the system model simulated by using the Euler-Maruyama scheme. The transition time between a high amplitude response and the low amplitude response is highlighted with the dashed line. }
\label{OneCantilever_EucledianNumericalTrial}
\end{figure}

The escape time out of the high amplitude basin  of \eqref{eq_OneCantileverODE} is estimated by using numerical experiments. The Euler-Maruyama method \cite{higham_algorithmic_2001} is used to integrate the SDE forward in time with the high amplitude attractor as the initial condition. A small timestep of $0.0001$ is used for the simulation.
 In Fig. \ref{OneCantilever_EucledianNumericalTrial}, the Euclidean norm of the Poincare section of a single simulation starting at the high amplitude attractor and transitioning to the low amplitude attractor due to the influence of noise with multiplier, $\sigma = 0.0045$ is shown.

\begin{figure}[htbp]
\centerline{\includegraphics[width=0.5\textwidth]{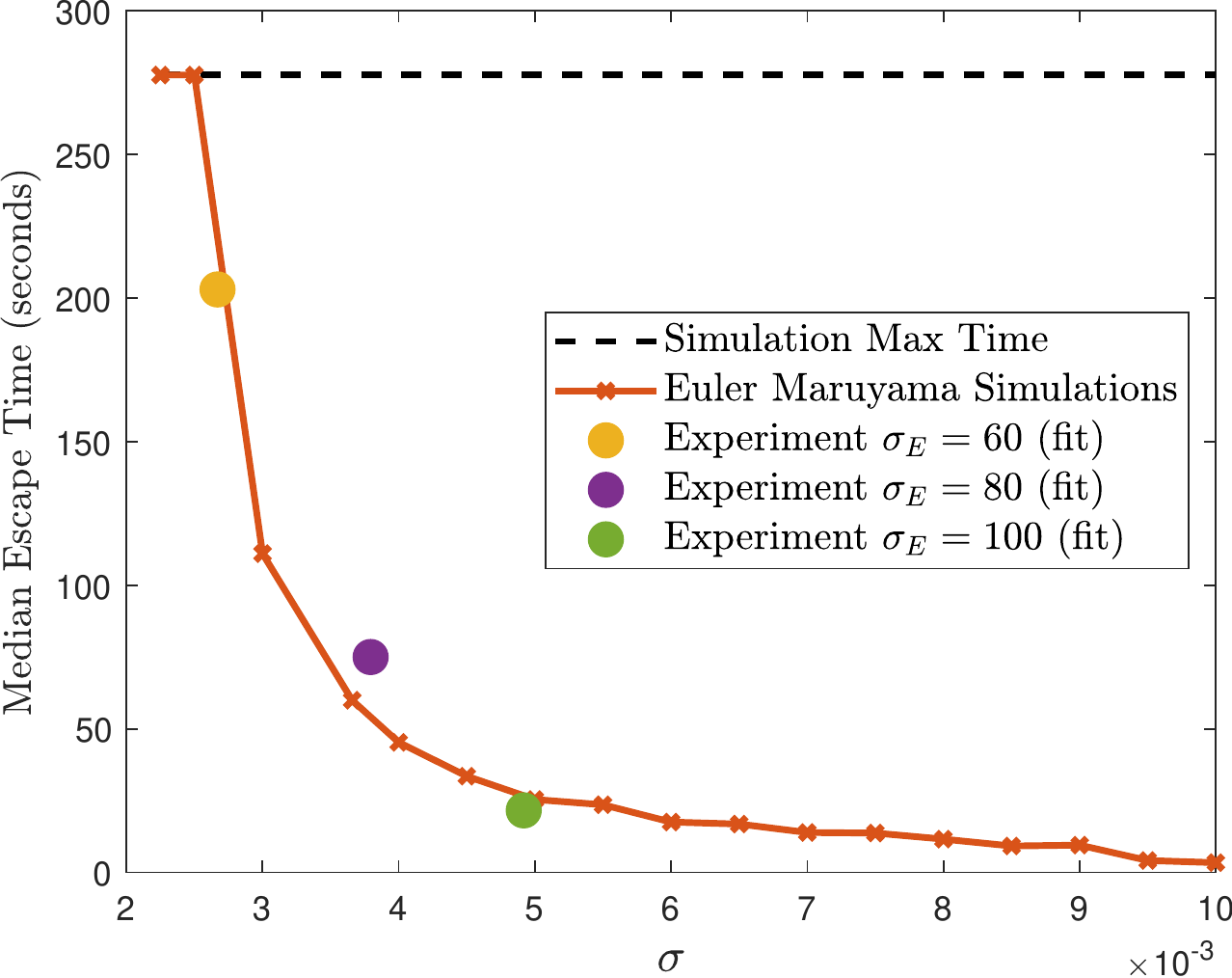}}
\caption{The median escape time for sets of Euler-Maruyama simulations that begin at the high amplitude attractor as a function of the noise intensity $\sigma$. Each data point is calculated by using $200$ Euler-Maruyama simulations. For each simulation, $\Omega = 0.9$ that is equivalent to $f_i = 18.00 $Hz. The experimental escape times are fit to the natural log simulated escape times for comparison.}
\label{OneCantilever_MedianEscapeTimeEM}
\end{figure}

For each $\sigma$, 200 trials of the numerical experiment are conducted. Each trial had a fixed end time of 5000 periods of excitation. These are long duration trials; each numerical trial has $3.5\cdot10^{8}$ timesteps, and takes on the average about $50$ minutes to compute. High performance parallelization of the numerical trials is performed on the Maryland Advanced Research Computing Center.  
The median escape times recorded from the numerical experiments are shown in Fig. \ref {OneCantilever_MedianEscapeTimeEM}.

\begin{figure}[htbp]
\centerline{\includegraphics[width=0.5\textwidth]{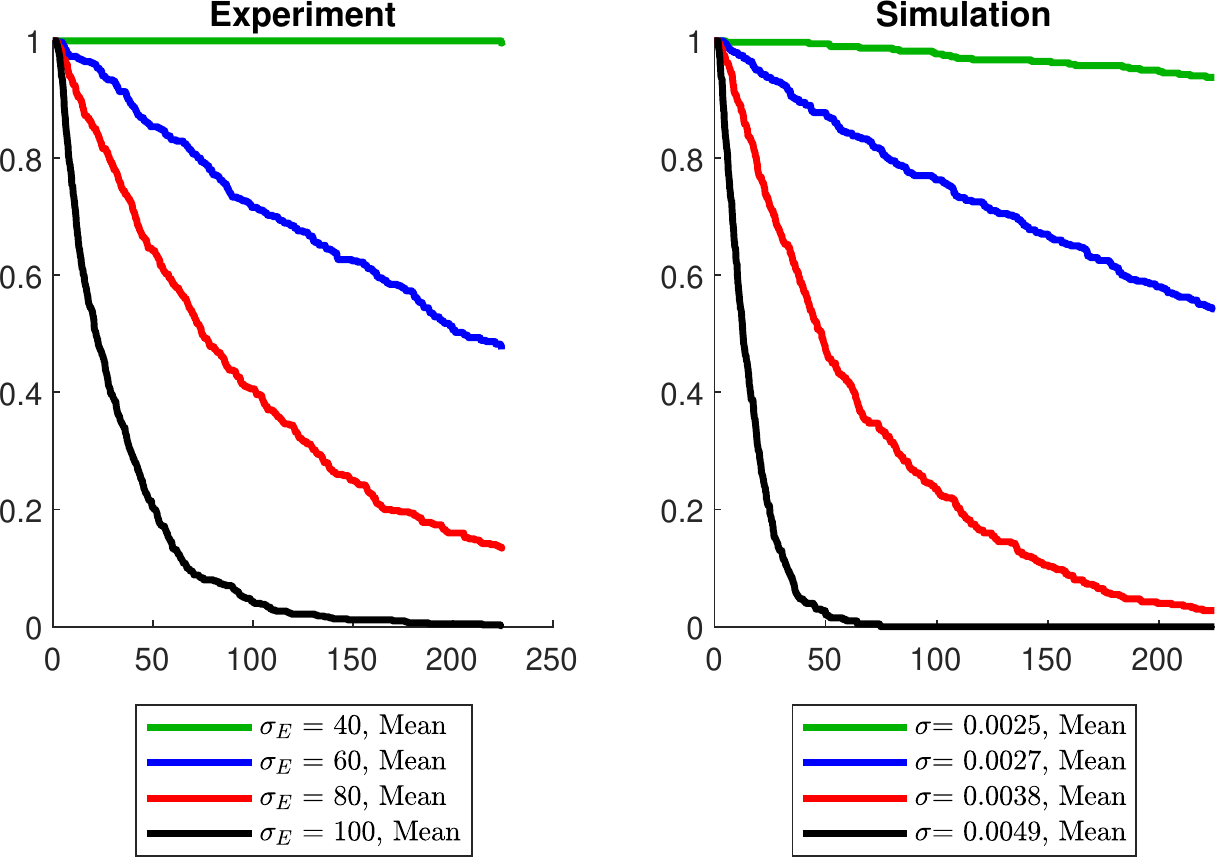}}
\caption{Estimates of the probability of remaining in the high amplitude attractor of the experiment (left) and simulation (right) data as a function of time and noise intensity. Here, $f_i = 18$Hz.}
\label{OneCantilever_NumericalProbabilityEstimates18Hz}
\end{figure}

For $\sigma < 3\cdot10^{-3}$, the median escape time exceeds the maximum time of the numerical trials. For larger noise intensities, the simulations follow an exponential decay of median escape times. In Fig. \ref {OneCantilever_MedianEscapeTimeEM}, the experimental escape times are graphed with the numerical escape times. Note that $\sigma$, the noise multiplier from \eqref{eq_OneCantileverSDE}, is not directly comparable to $\sigma_E$, the experimental noise intensity. To create Fig. \ref {OneCantilever_MedianEscapeTimeEM}, least squares optimization is performed to minimize the error between the escape times in the natural log scale. Resulting from the optimization, $\sigma_E$ is rescaled to fit the numerical experiment. There is high sensitivity of the escape times to noise intensity, and these fitted comparisons are performed to show only qualitative similarities between the experimental and numerical results. The experimental noise intensities of $\sigma_E = 60,$ $80,$ and $100$ correspond to  $\sigma = 0.0027,$ $0.0038,$ and $0.0049$, respectively after fitting.

The probability estimates for \eqref{eq_OneCantileverSDE} to remain in the basin of the high amplitude attractor at these noise intensities are shown in Fig. \ref {OneCantilever_NumericalProbabilityEstimates18Hz}. These estimates are simulated and computed for the harmonic excitation frequency of $18.00 Hz$. The resulting trend of decreasing probability with higher noise intensities is similar to what was observed with the experimental system. 


\begin{figure}[htbp]
\centerline{\includegraphics[width=0.5\textwidth]{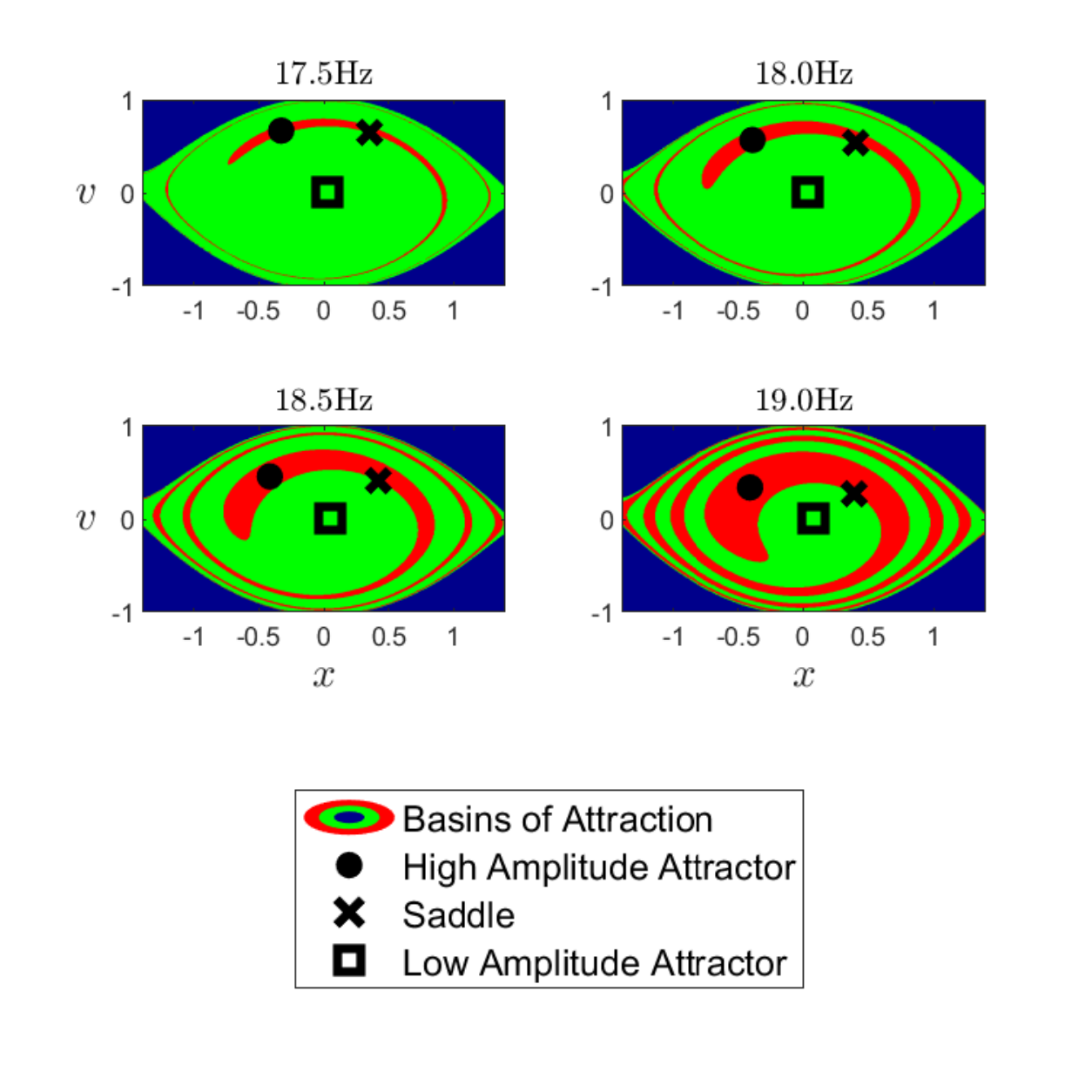}}
\caption{Basins of attraction of the system model for different excitation frequencies, $f_i$. Notice that the domain of the high amplitude basin of attraction increases in area with the excitation frequency. }
\label{OneCantilever_BasinsFrequencies}
\end{figure}

An exponential relationship between the excitation frequency and the escape time is observed in the experimental system. To  understand the origin of this characteristic, a study of the stochastic behavior of the system model  \label{eq_OneCantileverSDE} at various excitation frequencies is conducted. The basins of attraction of the system model at the frequencies $f_i = [17.50,18.00,18.50,19.00]$ Hz are shown in Fig. \ref {OneCantilever_BasinsFrequencies}.

\begin{figure}[htbp]
\centerline{\includegraphics[width=0.5\textwidth]{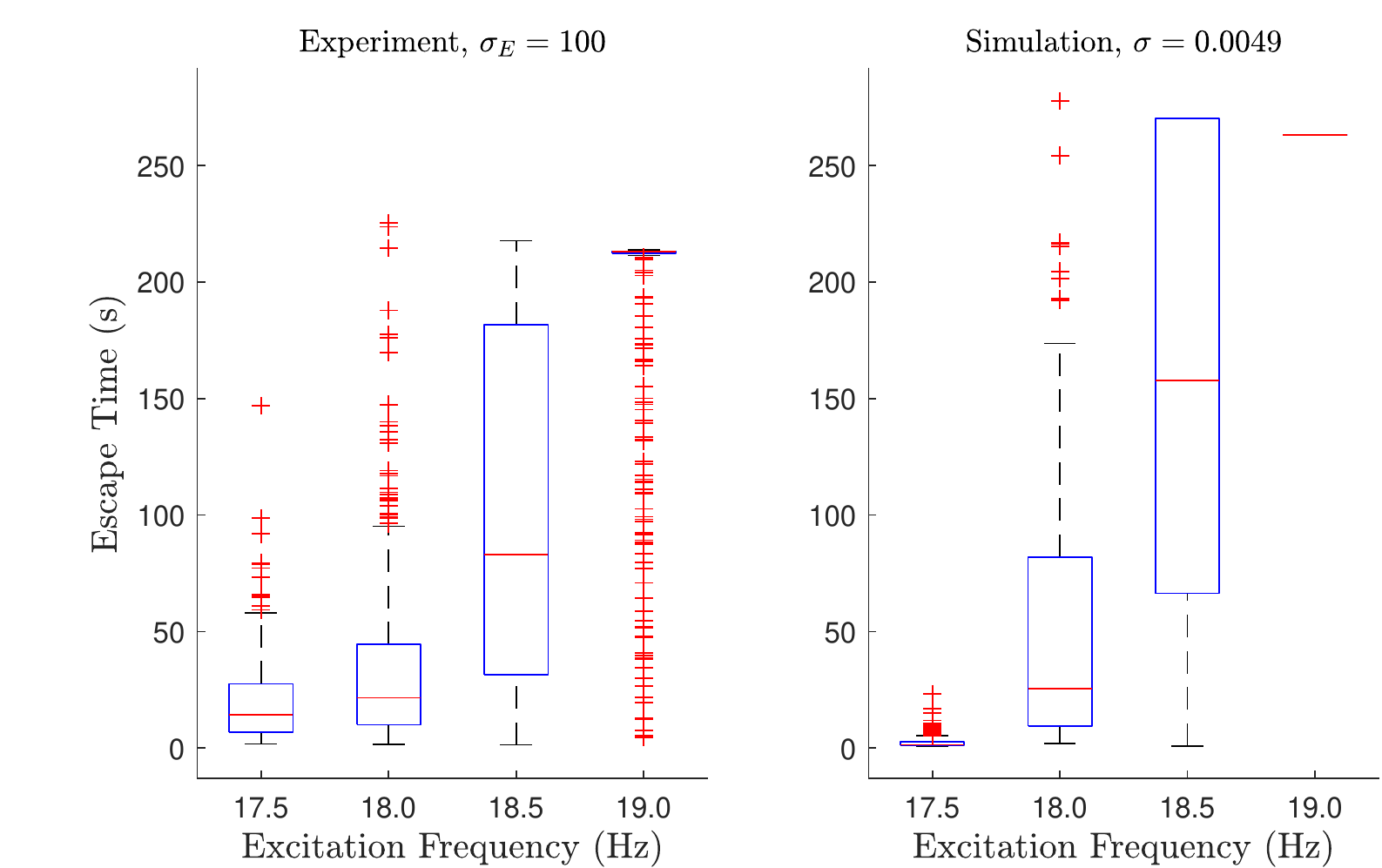}}
\caption{Box plots of the escape times of the experiment (left) and the Euler-Maruyama simulations (right) as a function of excitation frequency.}
\label{OneCantilever_MedianEscapeTimeEMFrequencies}
\end{figure}

 Evidently, the high amplitude basin of attraction shown in red increases in size  as the excitation frequency grows. 
Box plots of the escape times as a function of excitation frequency are shown in Fig. \ref {OneCantilever_MedianEscapeTimeEMFrequencies}. Similar to the experimental trials, the simulations results indicate an exponential relationship  between the excitation frequency and the median escape time.

\begin{figure}[htbp]
\centerline{\includegraphics[width=0.5\textwidth]{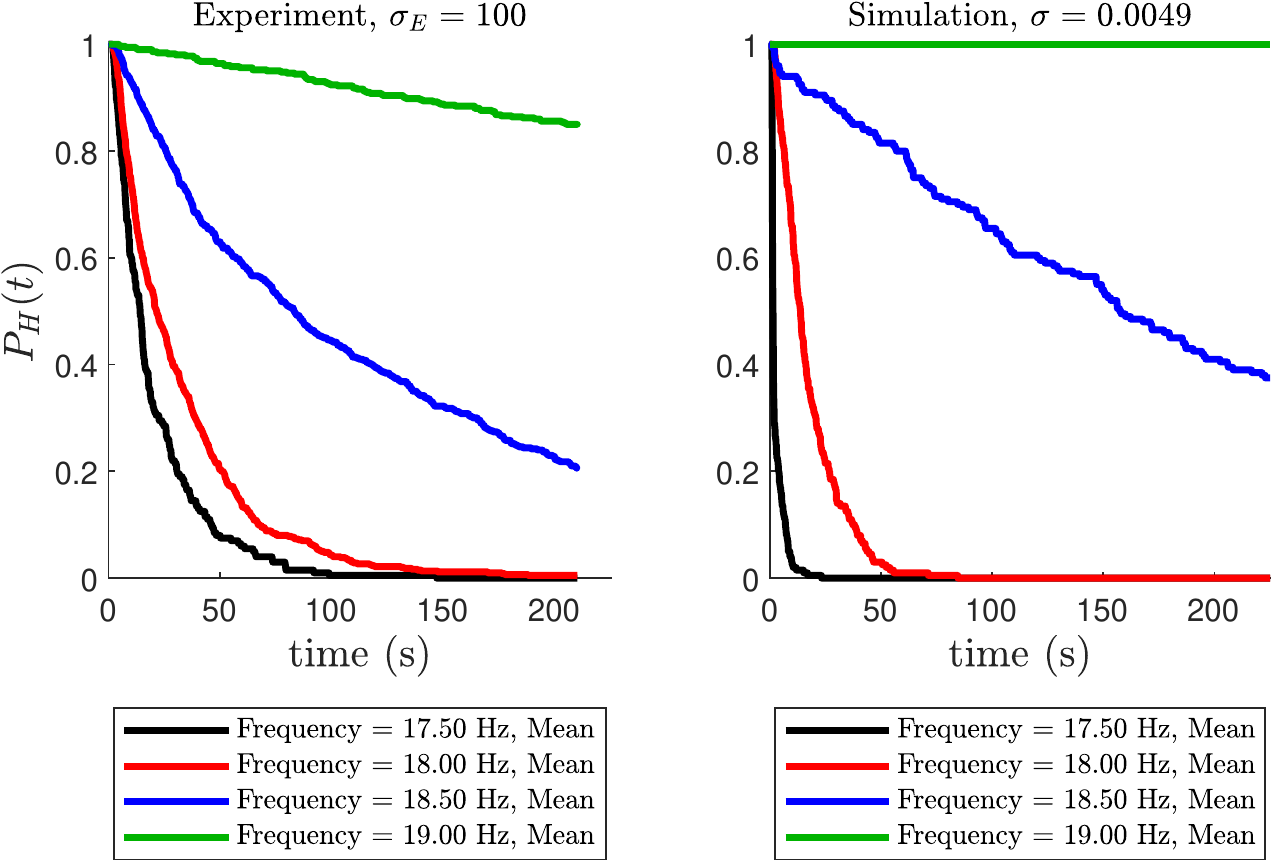}}
\caption{Estimates of the probability of remaining in the high amplitude attractor of the experiment (left) and simulation (right) data as a function of time and frequency. }
\label{OneCantilever_NumericalProbabilityEstimatesFrequencies}
\end{figure}

The probabilities of remaining in the high amplitude basin as a function of time and excitation frequency are shown in Fig. \ref {OneCantilever_NumericalProbabilityEstimatesFrequencies}. The results in this figure are qualitatively comparable to the experimental results in Fig. \ref {OneCantilever_ExperimentalProbabilityEstimateN100}.


\section{ Quasipotential analysis} \label{Section_QuasipotentialAnalysis}

In this section, the quasipotential barriers are estimated by (i) fitting to the results of Monte Carlo simulations and 
 (ii) using the approach developed in earlier work \cite{cilenti_most_2022}.
 The quasipotential, $U$ is a quantitative measure of the difficulty of escape so that the expected escape time is proportional to $\exp{-U/\sigma^2}$. Formal definitions for $U$ from the Freidlin and Wentzell are available in  prior work \cite{freidlin_random_1998}. 
 
\begin{figure}[htbp]
\centerline{\includegraphics[width=0.5\textwidth]{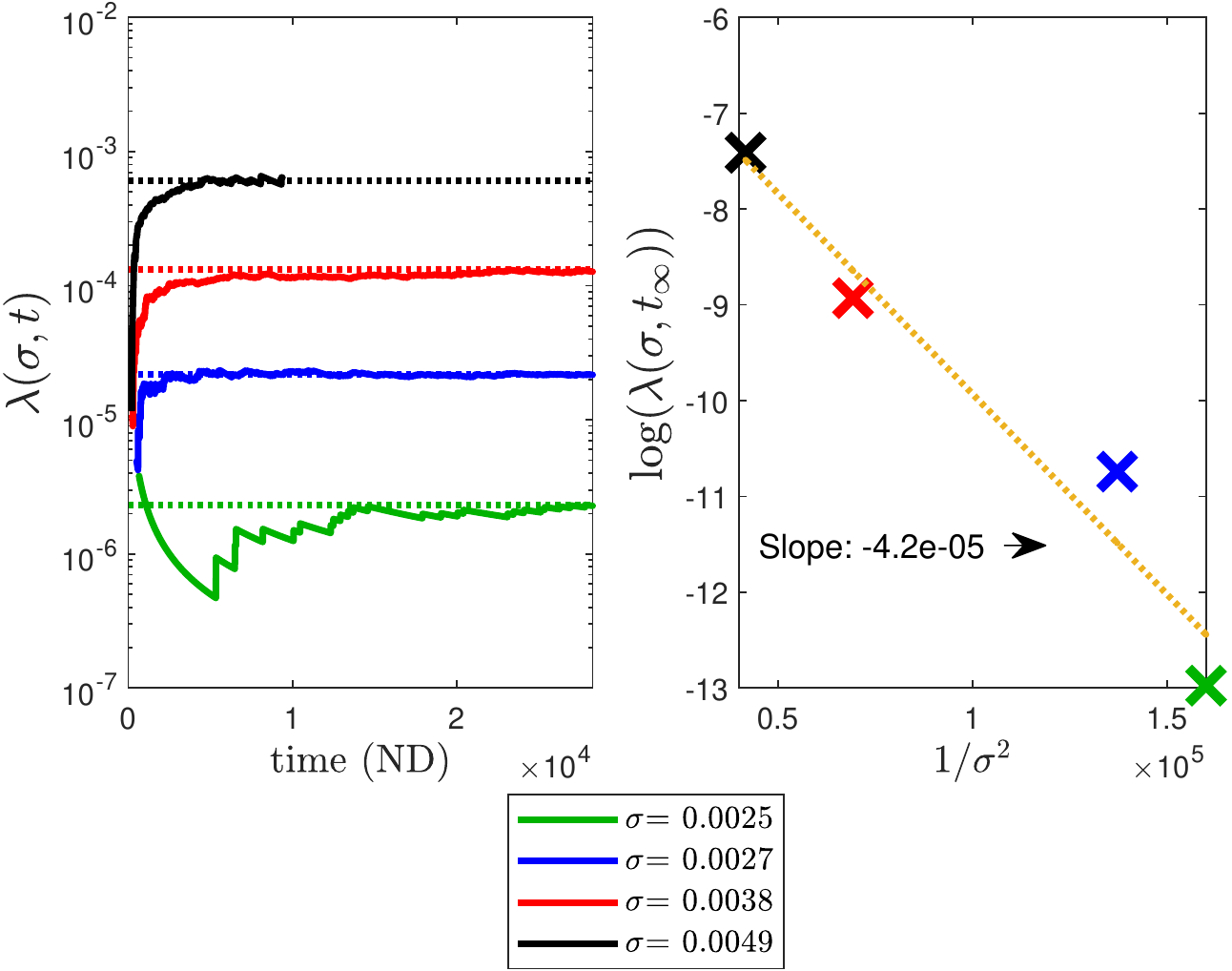}}
\caption{Estimates of the quasipotential $U$ extracted from the slope of $\log(\lambda(\sigma))$ and $1/\sigma^2$.}
\label{OneCantilever_QuasipotentialEstimates18Hz}
\end{figure}

 The probability of remaining in the high amplitude basin is assumed to have the following model:
\begin{equation}
    \begin{split}
         P(\tau_{escape} >t) & = \exp{ -\lambda(\sigma) t},  \\
        {\rm where}~~ \lambda(\sigma) & = C \exp{-U/\sigma^2}.
    \end{split}
\end{equation}
Then 
\begin{align*}
\lambda(\sigma)& = -\frac{1}{t} \log P(\tau_{escape} > t)\quad {\rm and} \\
\log(\lambda(\sigma))&= -\frac{U}{\sigma^2} + \log C.
\end{align*}
 The escape rates $\lambda(\sigma,t)$ for various noise intensities are calculated at all times $0<t<5000T$ where $T = 2\pi/\Omega$ is the period. The results for $\Omega = 0.90$ that corresponds to the excitation frequency of 18.00 Hz are displayed in Fig. \ref {OneCantilever_QuasipotentialEstimates18Hz}. As expected, $\lambda(\sigma,t)$ approach constant values as $t\rightarrow\infty$. 

 The probability curves in Fig. \ref {OneCantilever_NumericalProbabilityEstimates18Hz} can be used to estimate the quasipotential.
Assuming that $P(\tau_{escape} >t) \approx P_H(t)$, the quasipotential $U$ can be estimated as the slope of the best fit line between $\log(\lambda(\sigma))$ and $1/\sigma^2$ as shown in Fig. \ref {OneCantilever_QuasipotentialEstimates18Hz}.
The negative of the slope of the best fit line to $\log(\lambda(\sigma))$ and  $1/\sigma^2$ yields $U \approx 4.2\cdot10^{-5}$.

\begin{figure}[htbp]
\centerline{\includegraphics[width=0.5\textwidth]{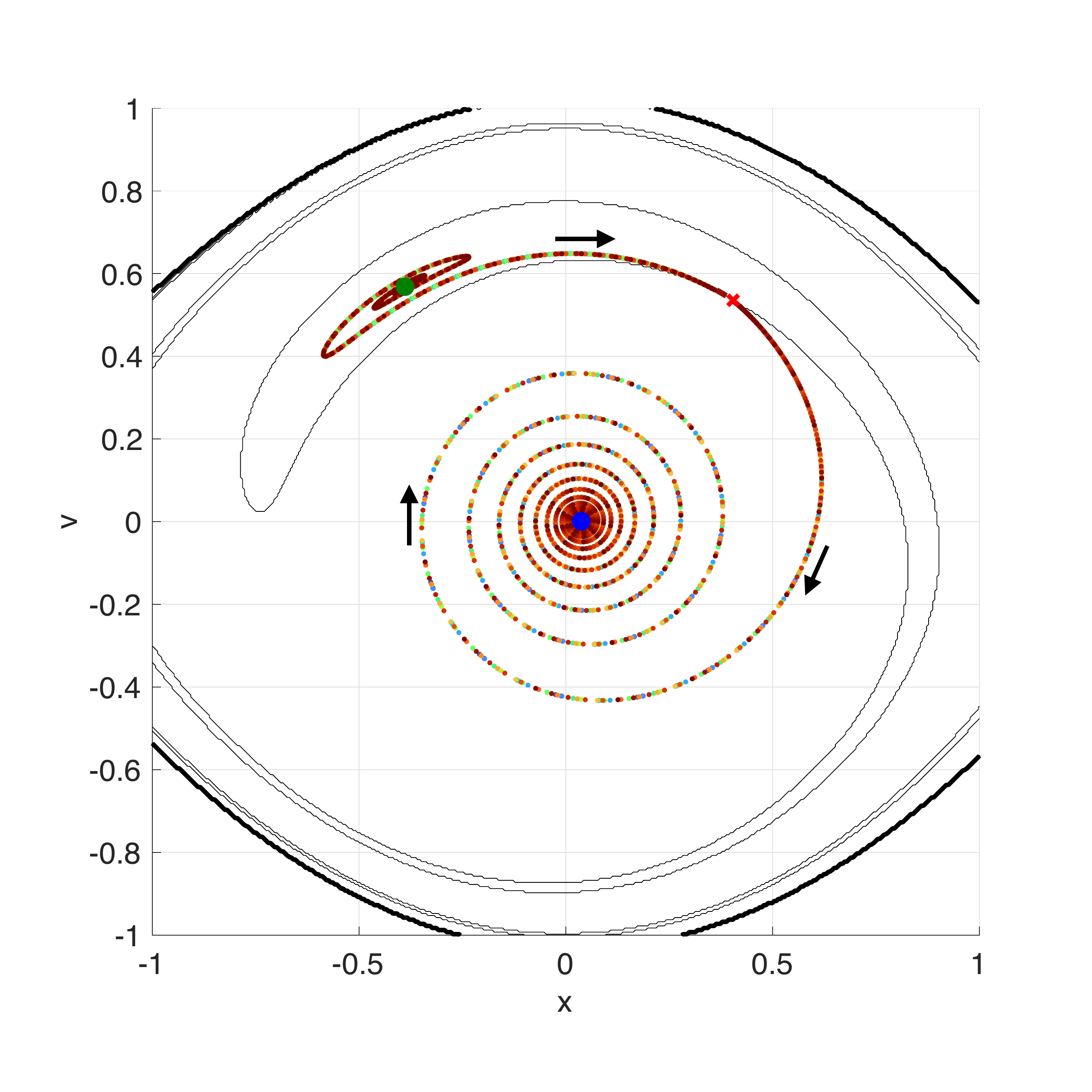}}
\caption{Most probable escape paths from the high attractor to the saddle and an ensemble of trajectories from the saddle to the low attractor.}
\label{TransitionPathsHighLow}
\end{figure}

\begin{figure}[htbp]
\centerline{\includegraphics[width=0.5\textwidth]{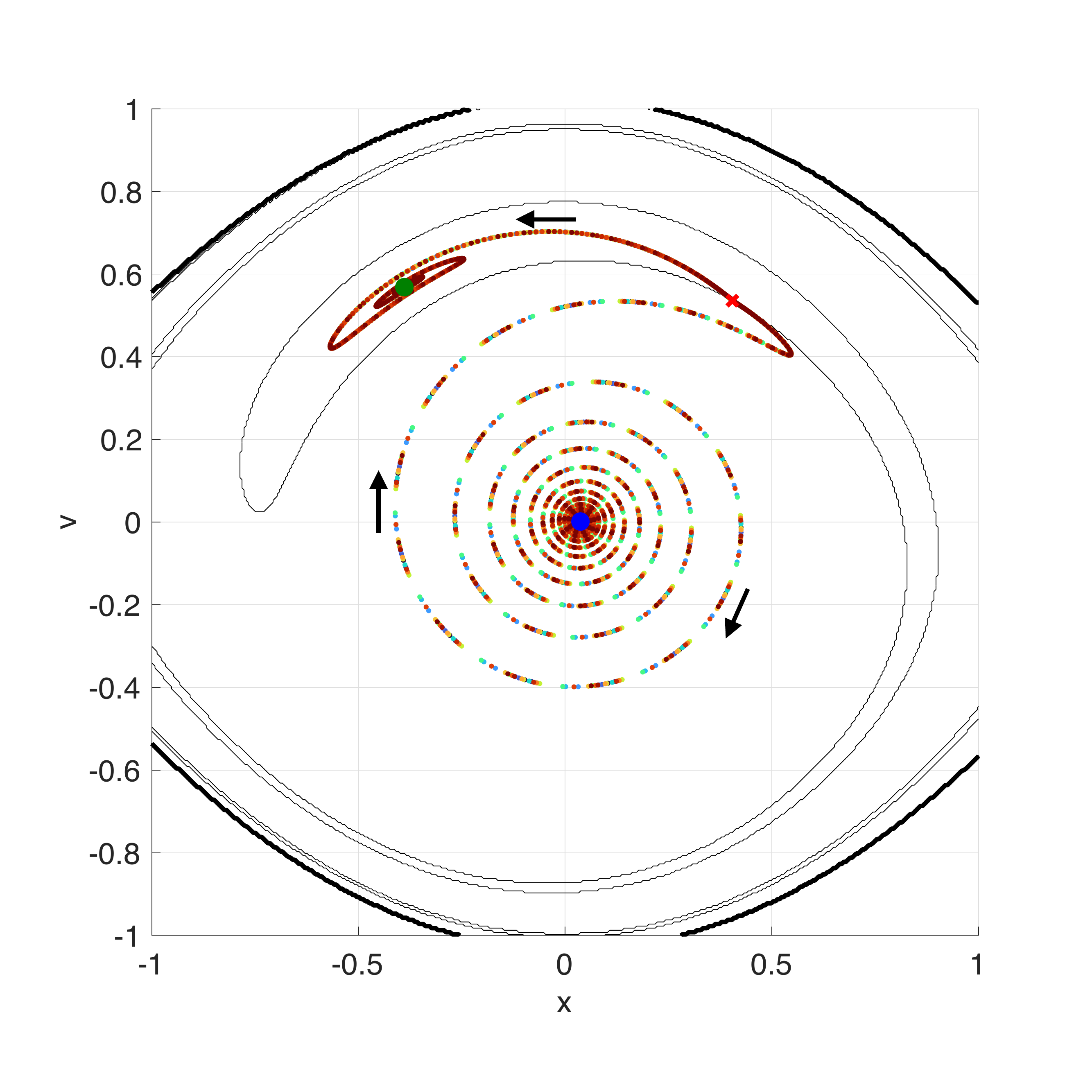}}
\caption{Most probable escape paths from the low attractor to the saddle and an ensemble of trajectories from the saddle to the high attractor.}
\label{TransitionPathsLowHigh}
\end{figure}

 The quasipotential was also computed for this system at different excitation frequencies by using the deterministic methodology in earlier work \cite{cilenti_most_2022}. At the excitation frequency of $\Omega = 0.9$, the quasipotential found using the deterministic methodology is $4.21\cdot10^{-5}$. This quasipotential corresponds to the minimum action path that transitions from the high amplitude to the low amplitude attractor as shown in Fig. \ref {TransitionPathsHighLow}. There is less than $<0.1\%$  difference between the stochastic estimate of the quasipotential and the deterministic estimate, which is quite remarkable. The quasipotential barrier for the escape from the low attractor to the high attractor is found to be $2.82\cdot 10^{-3}$. This quasipotential corresponds to the minimum action path that transitions from the low amplitude attractor to the high amplitude attractor as shown in \ref{TransitionPathsLowHigh} and is larger than the barrier for the escape from the high attractor to the low attractor by a factor of about 67. 

\begin{figure}[htbp]
\centerline{\includegraphics[width=0.5\textwidth]{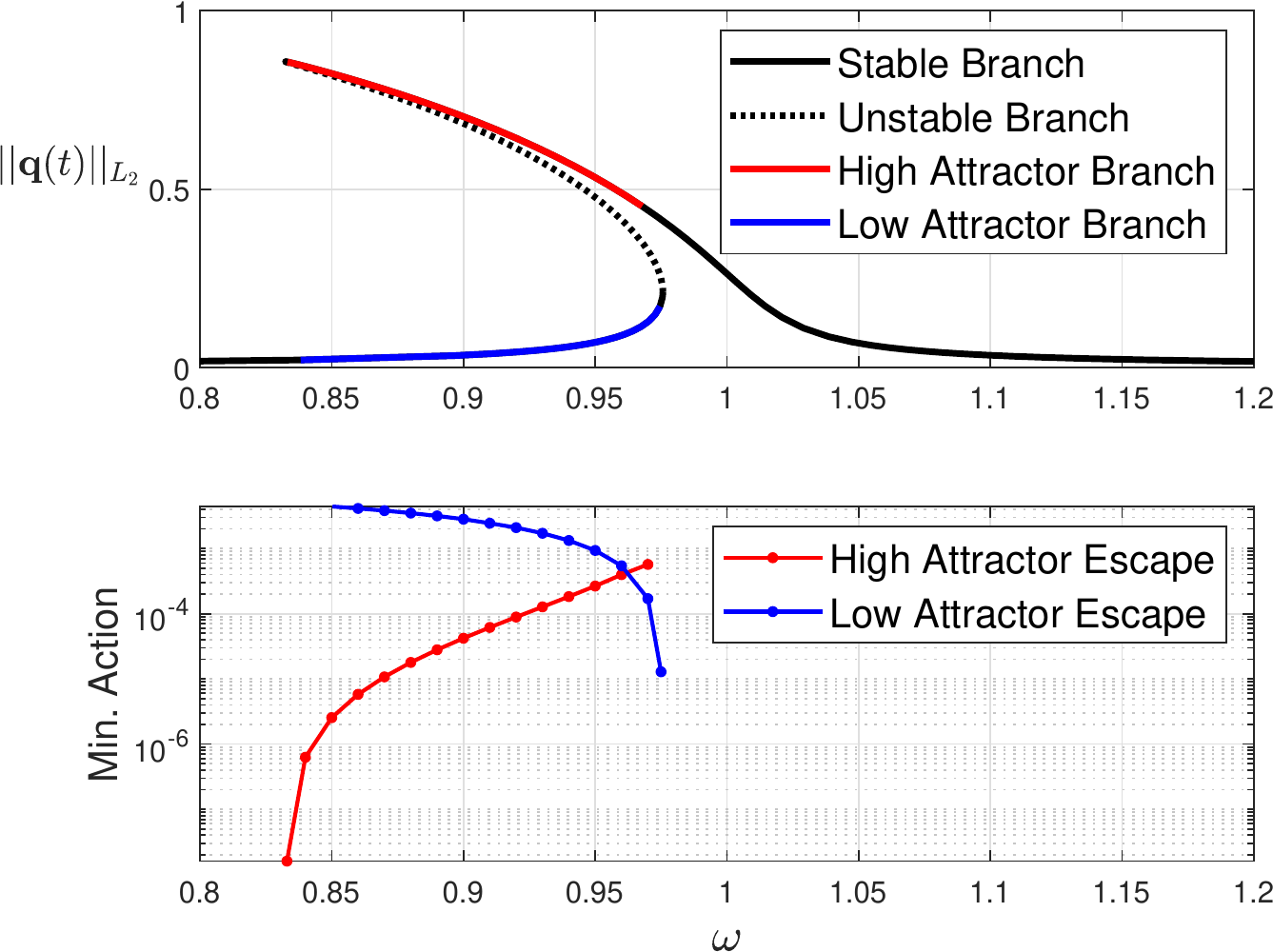}}
\caption{The quasipotential as a function of frequency for escapes out of the high amplitude and low amplitude attractors, shown in red and blue, respectively. Notice that the quasipotential to escape the high amplitude attractor is lower in magnitude than the quasipotential to escape the low amplitude attractor for the majorities, except for those very close to the jump up frequency. This suggests the probability of transitioning from high amplitude response to low amplitude response far exceeds the probability of transitioning from low amplitude response to high amplitude response for the majority of frequencies of excitation.   }
\label{OneCantilever_FrequencyQuasipotential}
\end{figure}

 The quasipotential for different excitation frequencies of \eqref{eq_OneCantileverSDE} are shown in Fig. \ref {OneCantilever_FrequencyQuasipotential}.

It is remarked  that the minimum action to escape the high amplitude attractor is much smaller than the minimum action to escape the low amplitude attractor, except near the jump up frequencies. The dependence of the quasipotential on the direction of the transition and on the excitation frequency explains the observed phenomenon that no transitions from the low amplitude to high amplitude responses were observed in the experiments.


\section{Discussion}
\label{Section_OneCantileverDiscussion}

In this work, the effect of noise on a rotating cantilever structure and on a softening Duffing oscillator were examined. The experimental system was shown to demonstrate characteristic nonlinear phenomena, including multiple stable responses under the influence of an external periodic excitation. The experiment was designed to produce large number of long duration noise trials. Through these noise trials, evidence was created to show that noise could  cause transitions from the high amplitude response to the low amplitude response, with probabilities that depend on the intensity of the noise and the frequency of the external excitation.

In addition, hundreds of long duration noise trials were run that started at the low amplitude response. These trials were performed with a variety of noise intensities and excitation frequencies. The system did not transition from the low amplitude response to the high amplitude response in any of these trials. That is, it is so much less probable for this transition to occur that it was never observed experimentally in the considered range of frequencies. There was an exception to this observation: transitions did occur to a higher amplitude response at the jump up frequencies, however, these transitions when they were observed occurred with or without noise being added to the control signal. 

By itself, the experimental results can provide interesting new insights. The cantilever beam structure could, for example, represent a blade of turbomachinery. It was demonstrated that if this beam experiences nonlinear restoring forces, it can exhibit potentially undesired high amplitude responses. Furthermore, it was demonstrated that for certain designs influenced by noise, with much higher probability, the system will transition out of the high amplitude response than into the high amplitude response.  This nearly one directional behavior supports the idea that noise from the environment or an open loop control strategy can be useful for keeping a system
in desirable operating conditions. 

In order to explain the source of the stochastic nonlinear behavior in the experiment, a nonlinear, single degree-of-freedom discrete mass, spring-mass-damper system model for the cantilever structure was proposed. This model is a significant simplification of the experiment, yet, this model is useful to gain insights into the system behavior. 
\begin{itemize}
\item It was demonstrated that the Euler-Maruyama simulations of the system model qualitatively reproduce the relationship between escape times and probability of escape with noise intensity and excitation frequency parameters. 
\item Furthermore, it was demonstrated that similar transition paths can be observed in the experiments. 
\item Additionally, it was shown how the quasipotential varies with the excitation frequency of this system; it was observed that the quasipotential of the escape from the high amplitude response are much smaller than the quasipotential of escapes from the low amplitude response. 
\end{itemize}
It stands to reason that the observed nearly one directional transition behavior in the experiments occurs because the amount of action (work) required to escape the low amplitude response is much greater than the amount of action (work) to escape the high amplitude response  at the excitation frequencies studied experimentally. Moreover, it was found that for the system model, the quasipotential barrier for the escape from the high-amplitude to the low-amplitude attractor is lower than that for the escape from the low-amplitude to the high-amplitude attractor throughout most of the range of frequencies where bistability takes place.  

The nonlinear system model with numerical tools such as the one used in this work to quantify the probability of transitions and the action required for transitions could be used by designers to develop physical nonlinear systems that exhibit desirable responses to noise. 

The system model has limitations. The best fit of the system model to the experimental data diverges from the response of the model. Furthermore, the noise multiplier in the experiment could not be compared via derivation to the noise multiplier in the system model. This is in part because white noise is assumed in the system model, while filtered white noise is delivered by the actuator in the experiments.  Furthermore, the observed behavior in the experiments and with the system model are found to be highly sensitive to noise intensity, and small errors from assumptions in a derivation could lead to significantly different expected results. 

Comparisons were created between the simulations and the experiments by using a best fit data driven approach between the measured escape times and the noise intensities. The resulting comparisons have errors that are adequate for qualitative observations of how the systems respond to parameter changes, but may be inadequate for prediction of the stochastic response. Note that the latter is not an objective of this work. Considerations of additional nonlinear terms, a model with more degrees of freedom, filtered noise models, or noise models with noise intensities calculated from the power spectrum could improve on the comparisons  and be a subject of future work.  

\section*{Acknowledgments}
This work was partially supported by  NSF grant CMMI-1760366 and associated data science supplements (BB) and by AFOSR MURI grant FA9550-20-1-0397 (MC). A part of this research project was conducted by using computational resources available at the Maryland Advanced Research Computing Center (MARCC).



 \bibliographystyle{elsarticle-num}

\begin{thebibliography}{10}
\expandafter\ifx\csname url\endcsname\relax
  \def\url#1{\texttt{#1}}\fi
\expandafter\ifx\csname urlprefix\endcsname\relax\def\urlprefix{URL }\fi
\expandafter\ifx\csname href\endcsname\relax
  \def\href#1#2{#2} \def\path#1{#1}\fi

\bibitem{grolet_free_2012}
A.~Grolet, F.~Thouverez,
  \href{http://www.sciencedirect.com/science/article/pii/S0022460X12001174}{Free
  and forced vibration analysis of a nonlinear system with cyclic symmetry:
  {Application} to a simplified model}, Journal of Sound and Vibration 331~(12)
  (2012) 2911--2928.
\newblock \href {http://dx.doi.org/10.1016/j.jsv.2012.02.008}
  {\path{doi:10.1016/j.jsv.2012.02.008}}.
\newline\urlprefix\url{http://www.sciencedirect.com/science/article/pii/S0022460X12001174}

\bibitem{balachandran_response_2015}
B.~Balachandran, E.~Perkins, T.~Fitzgerald, Response localization in
  micro-scale oscillator arrays: influence of cubic coupling nonlinearities,
  International Journal of Dynamics and Control 3~(2) (2015) 183--188,
  publisher: Springer.

\bibitem{papangelo_multistability_2019}
A.~Papangelo, F.~Fontanela, A.~Grolet, M.~Ciavarella, N.~Hoffmann,
  \href{http://www.sciencedirect.com/science/article/pii/S0022460X18307016}{Multistability
  and localization in forced cyclic symmetric structures modelled by
  weakly-coupled {Duffing} oscillators}, Journal of Sound and Vibration 440
  (2019) 202--211.
\newblock \href {http://dx.doi.org/10.1016/j.jsv.2018.10.028}
  {\path{doi:10.1016/j.jsv.2018.10.028}}.
\newline\urlprefix\url{http://www.sciencedirect.com/science/article/pii/S0022460X18307016}

\bibitem{bartels_computational_2007}
R.~E. Bartels, A.~Sayma, Computational aeroelastic modelling of airframes and
  turbomachinery: progress and challenges, Philosophical Transactions of the
  Royal Society A: Mathematical, Physical and Engineering Sciences 365~(1859)
  (2007) 2469--2499, publisher: The Royal Society London.

\bibitem{jia_review_2020}
Y.~Jia, Review of nonlinear vibration energy harvesting: {Duffing},
  bistability, parametric, stochastic and others, Journal of Intelligent
  Material Systems and Structures 31~(7) (2020) 921--944, publisher: SAGE
  Publications Sage UK: London, England.

\bibitem{agarwal_influence_2018}
V.~Agarwal, X.~Zheng, B.~Balachandran, Influence of noise on frequency
  responses of softening {Duffing} oscillators, Physics Letters A 382~(46)
  (2018) 3355--3364, iSBN: 0375-9601 Publisher: Elsevier.

\bibitem{perkins_effects_2016}
E.~Perkins, M.~Kimura, T.~Hikihara, B.~Balachandran, Effects of noise on
  symmetric intrinsic localized modes, Nonlinear Dynamics 85~(1) (2016)
  333--341, iSBN: 0924-090X Publisher: Springer.

\bibitem{alofi_noise_2022}
A.~Alofi, G.~Acar, B.~Balachandran,
  \href{https://www.sciencedirect.com/science/article/pii/S0022460X22001808}{Noise
  influenced response movement in coupled oscillator arrays with
  multi-stability}, Journal of Sound and Vibration (2022) 116951\href
  {http://dx.doi.org/10.1016/j.jsv.2022.116951}
  {\path{doi:10.1016/j.jsv.2022.116951}}.
\newline\urlprefix\url{https://www.sciencedirect.com/science/article/pii/S0022460X22001808}

\bibitem{perkins_noise-influenced_2013}
E.~Perkins, C.~Chabalko, B.~Balachandran, Noise-influenced transient energy
  localization in an oscillator array, Nonlinear Theory and Its Applications,
  IEICE 4~(3) (2013) 232--243, iSBN: 2185-4106 Publisher: The Institute of
  Electronics, Information and Communication Engineers.

\bibitem{perkins_noise-influenced_2015}
J.~E. Perkins, Noise-influenced dynamics of nonlinear oscillators, {PhD}
  {Thesis}, University of Maryland, College Park (2015).

\bibitem{haitao_dynamics_2015}
L.~Haitao, Q.~Weiyang, L.~Chunbo, D.~Wangzheng, Z.~Zhiyong,
  \href{http://dx.doi.org/10.1088/0964-1726/25/1/015001}{Dynamics and coherence
  resonance of tri-stable energy harvesting system}, Smart Materials and
  Structures 25~(1) (2015) 015001, publisher: IOP Publishing.
\newblock \href {http://dx.doi.org/10.1088/0964-1726/25/1/015001}
  {\path{doi:10.1088/0964-1726/25/1/015001}}.
\newline\urlprefix\url{http://dx.doi.org/10.1088/0964-1726/25/1/015001}

\bibitem{balachandran_dynamics_2022}
B.~Balachandran, T.~Breunung, G.~D. Acar, A.~Alofi, J.~A. Yorke, Dynamics of
  circular oscillator arrays subjected to noise, Nonlinear Dynamics (2022)
  1--14Publisher: Springer.

\bibitem{duffing_erzwungene_1918}
G.~Duffing, Erzwungene schwingungen bei veränderlicher eigenfrequenz, Vieweg
  u. Sohn, Braunschweig 7.

\bibitem{lingala2017random}
N.~Lingala, N.~S. Namachchivaya, I.~Pavlyukevich, Random perturbations of a
  periodically driven nonlinear oscillator: escape from a resonance zone,
  Nonlinearity 30~(4) (2017) 1376.

\bibitem{ren_local_2019}
Z.~Ren, W.~Xu, Y.~Qiao, Local averaged path integration method approach for
  nonlinear dynamic systems, Applied Mathematics and Computation 344 (2019)
  68--77, publisher: Elsevier.

\bibitem{agarwal_noise-induced_2020}
V.~Agarwal, J.~A. Yorke, B.~Balachandran, Noise-induced chaotic-attractor
  escape route, Nonlinear Dynamics 102~(2) (2020) 863--876, publisher:
  Springer.

\bibitem{kikuchi_ritz_2020}
L.~Kikuchi, R.~Singh, M.~E. Cates, R.~Adhikari,
  \href{https://link.aps.org/doi/10.1103/PhysRevResearch.2.033208}{Ritz method
  for transition paths and quasipotentials of rare diffusive events}, Phys.
  Rev. Research 2~(3) (2020) 033208, publisher: American Physical Society.
\newblock \href {http://dx.doi.org/10.1103/PhysRevResearch.2.033208}
  {\path{doi:10.1103/PhysRevResearch.2.033208}}.
\newline\urlprefix\url{https://link.aps.org/doi/10.1103/PhysRevResearch.2.033208}

\bibitem{chao_tao_2021}
Y.~Chao, M.~Tao, Parametric resonance for enhancing the rate of metastable
  transition.

\bibitem{cilenti_transient_2021}
L.~Cilenti, B.~Balachandran, \href{https://doi.org/10.1063/5.0051103}{Transient
  probability in basins of noise influenced responses of mono and coupled
  duffing oscillators} 31~(6)  063117, publisher: American Institute of
  Physics.
\newblock \href {http://dx.doi.org/10.1063/5.0051103}
  {\path{doi:10.1063/5.0051103}}.
\newline\urlprefix\url{https://doi.org/10.1063/5.0051103}

\bibitem{zhang_koopman_2021}
B.~Zhang, T.~Sahai, Y.~Marzouk, A {Koopman} framework for rare event simulation
  in stochastic differential equations, arXiv preprint arXiv:2101.07330.

\bibitem{cilenti_most_2022}
L.~Cilenti, M.~Cameron, B.~Balachandran,
  \href{https://doi.org/10.1063/5.0093074}{Most probable escape paths in
  periodically driven nonlinear oscillators}, Chaos: An Interdisciplinary
  Journal of Nonlinear Science 32~(8) (2022) 083140, publisher: American
  Institute of Physics.
\newblock \href {http://dx.doi.org/10.1063/5.0093074}
  {\path{doi:10.1063/5.0093074}}.
\newline\urlprefix\url{https://doi.org/10.1063/5.0093074}

\bibitem{kerschen_past_2006}
G.~Kerschen, K.~Worden, A.~F. Vakakis, J.-C. Golinval,
  \href{https://www.sciencedirect.com/science/article/pii/S0888327005000828}{Past,
  present and future of nonlinear system identification in structural
  dynamics}, Mechanical Systems and Signal Processing 20~(3) (2006) 505--592.
\newblock \href {http://dx.doi.org/10.1016/j.ymssp.2005.04.008}
  {\path{doi:10.1016/j.ymssp.2005.04.008}}.
\newline\urlprefix\url{https://www.sciencedirect.com/science/article/pii/S0888327005000828}

\bibitem{breunung_noise_2022}
T.~Breunung, B.~Balachandran,
  \href{https://www.sciencedirect.com/science/article/pii/S2095034922001003}{Noise
  {Color} {Influence} on {Escape} {Times} in {Nonlinear} {Oscillators} -
  {Experimental} and {Numerical} {Results}}, Theoretical and Applied Mechanics
  Letters (2022) 100420\href {http://dx.doi.org/10.1016/j.taml.2022.100420}
  {\path{doi:10.1016/j.taml.2022.100420}}.
\newline\urlprefix\url{https://www.sciencedirect.com/science/article/pii/S2095034922001003}

\bibitem{higham_algorithmic_2001}
D.~J. Higham, An algorithmic introduction to numerical simulation of stochastic
  differential equations, SIAM review 43~(3) (2001) 525--546, iSBN: 0036-1445
  Publisher: SIAM.

\bibitem{nayfeh_nonlinear_2008}
A.~Nayfeh, D.~Mook,
  \href{https://books.google.com/books?id=sj3ebg7jRaoC}{Nonlinear
  {Oscillations}}, Wiley {Classics} {Library}, Wiley, 2008.
\newline\urlprefix\url{https://books.google.com/books?id=sj3ebg7jRaoC}

\bibitem{dankowicz_coco_2019}
H.~Dankowicz, F.~Schilder, M.~Li, E.~Fotsch, M.~Henderson, {COCO}
  {Continuation} {Core} and {Toolboxes}, SourceForge, accessed Oct 17 (2019)
  2019.

\bibitem{freidlin_random_1998}
M.~I. Freidlin, A.~D. Wentzell, Random perturbations, in: Random perturbations
  of dynamical systems, Springer, 1998, pp. 15--43.

\end{thebibliography}

\end{document}